\documentclass[journal]{IEEEtran}
\ifCLASSINFOpdf
\else
\fi
\usepackage[utf8]{inputenc}
\usepackage{amsmath,amssymb,amsfonts}
\usepackage{cite}

\usepackage{algorithmic}
\usepackage{stfloats}
\usepackage{graphicx}
\usepackage{textcomp}
\usepackage{xcolor}
\usepackage{color}
\usepackage{pifont}
\usepackage{amsthm}
\newtheorem{theorem}{Theorem}

\usepackage{hyperref}
\usepackage{mathrsfs}
\usepackage{booktabs}
\usepackage{hyperref}
\usepackage{cases}
\usepackage{comment}
\usepackage{empheq} 
\usepackage{caption}
\usepackage{subcaption}
\usepackage{extarrows}
\allowdisplaybreaks[4]
\usepackage{url}  %Required
\usepackage{graphicx}  %Required
\usepackage[ruled,vlined,linesnumbered]{algorithm2e}

\usepackage{listings}

\begin{document}
\captionsetup[figure]{labelfont={rm},labelformat={default},labelsep=period,name={Fig.}}
% paper title
% Titles are generally capitalized except for words such as a, an, and, as,
% at, but, by, for, in, nor, of, on, or, the, to and up, which are usually
% not capitalized unless they are the first or last word of the title.
% Linebreaks \\ can be used within to get better formatting as desired.
% Do not put math or special symbols in the title.
\title{Movable Antenna-Aided Cooperative ISAC Network with {Time Synchronization} error and Imperfect CSI}
%
%
% author names and IEEE memberships
% note positions of commas and nonbreaking spaces ( ~ ) LaTeX will not break
% a structure at a ~ so this keeps an author's name from being broken across
% two lines.
% use \thanks{} to gain access to the first footnote area
% a separate \thanks must be used for each paragraph as LaTeX2e's \thanks
% was not built to handle multiple paragraphs
%

\author{Yue Xiu,~Yang Zhao,~Ran Yang,~Dusit Niyato,~\IEEEmembership{Fellow,~IEEE},~Jing Jin, Qixing Wang,~Guangyi Liu,~Ning Wei\\
%~Ning Wei,~\IEEEmembership{Member,~IEEE}\\
% <-this % stops a space
%\thanks{Yue Xiu, Songjie Yang, Ran Yang and Ning Wei are with 
%National Key Laboratory of Science and Technology on Communications, University of Electronic Science and Technology of China, Chengdu 611731, China (E-mail:  
%xiuyue12345678@163.com,yangsongjie@std.uestc.edu.cn,lyuwanting@yeah.net, wn@uestc.edu.cn).}
%\thanks{L. Qu is with the Faculty of Electrical Engineering
% and Computer Science of Ningbo University, 315211 Ningbo, China. (email: qulong@nbu.edu.cn)}
%\thanks{Chadi Assi is with Concordia University, Montreal, Quebec, H3G 1M8, Canada (E-mail:assi@ciise.concordia.ca).}
%\thanks{M. Debbah is with Khalifa University of Science and Technology, P O Box
%127788, Abu Dhabi, UAE (email: merouane.debbah@ku.ac.ae).
%.}
%\thanks{Yueyang Li is with School of Transportation and Logistics of Southwest Jiaotong University, 610032 Chengdu, China.(email: liyueyang@my.swjtu.edu.cn).
%.}
\thanks{The corresponding author is Ning Wei.}
}

\maketitle
% As a general rule, do not put math, special symbols or citations
% in the abstract or keywords.
\begin{abstract}
Cooperative-integrated sensing and communication (C-ISAC) networks have emerged as promising solutions for communication and target sensing. However, imperfect channel state information (CSI) estimation and time synchronization (TS) errors degrade performance, affecting communication and sensing accuracy. This paper addresses these challenges {by employing} {movable antennas} (MAs) to enhance C-ISAC robustness. We analyze the impact of CSI errors on achievable rates and introduce a hybrid Cramer-Rao lower bound (HCRLB) to evaluate the effect of TS errors on target localization accuracy. Based on these models, we derive the worst-case achievable rate and sensing precision under such errors. We optimize cooperative beamforming, {base station (BS)} selection factor and MA position to minimize power consumption while ensuring accuracy. {We then propose a} constrained deep reinforcement learning (C-DRL) approach to solve this non-convex optimization problem, using a modified deep deterministic policy gradient (DDPG) algorithm with a Wolpertinger architecture for efficient training under complex constraints. {Simulation results show that the proposed method significantly improves system robustness against CSI and TS errors, where robustness mean reliable data transmission under  poor channel conditions.}   These findings demonstrate the potential of MA technology to reduce power consumption in imperfect CSI and TS environments.
\end{abstract}

%近年来，协作集成感知和通信 (C-ISAC) 系统已成为一种有前途的解决方案，可同时支持通信和目标感知。这些系统依靠执行协调波束成形的分布式基站 (BS) 在进行静态目标感知的同时与用户通信。然而，实际应用中不完善的信道状态信息 (CSI) 估计和时间同步 (TS) 误差会导致性能下降，影响通信质量和感知精度。本文通过提出使用移动天线 (MA) 来增强 C-ISAC 系统的鲁棒性来解决这些挑战。具体来说，我们分析了 CSI 误差对可实现速率的影响，并引入了混合 Cramer-Rao 下限 (HCRLB) 来表征 TS 误差对目标定位精度的影响。使用这些模型，我们推导出在这种误差下最坏情况可实现的速率和感知精度。为了在确保目标感知精度和最坏情况可实现速率的同时最大限度地降低功耗，我们在 BS 设计了协作波束成形并优化了 MA 位置。为了解决由此产生的非凸优化问题，我们提出了一种约束深度强化学习 (C-DRL) 方法来动态优化波束成形、MA 定位和 BS 选择。具有 Wolpertinger 架构的改进深度确定性策略梯度 (DDPG) 算法可以在复杂约束和不同数量的用户下进行实际训练。仿真结果表明，所提出的方法显著提高了系统对 CSI 和 TS 误差的鲁棒性，在 MA 辅助 C-ISAC 场景中优于传统的固定天线系统。这些结果凸显了 MA 技术在不完美 CSI 和 TS 条件下降低功耗的潜力。
\begin{IEEEkeywords}
 Cooperative integrated sensing and communication, channel state information, time synchronization, movable antenna, constraint deep reinforcement learning. 
\end{IEEEkeywords}

% For peer review papers, you can put extra information on the cover
% page as needed:
% \ifCLASSOPTIONpeerreview
% \begin{center} \bfseries EDICS Category: 3-BBND \end{center}
% \fi
%
% For peerreview papers, this IEEEtran command inserts a page break and
% creates the second title. It will be ignored for other modes.
\IEEEpeerreviewmaketitle

\section{Introduction}

\IEEEPARstart{T}{he} development of sixth-generation (6G) wireless networks aims to support emerging applications like smart cities, autonomous driving, and intelligent manufacturing, requiring massive connectivity and high-precision sensing\cite{b1}. Integrated sensing and communication (ISAC) has become a key enabler for 6G, attracting significant interest from both academia and industry\cite{b2}. As wireless communication and sensing systems evolve, they are expected to operate in high-frequency broadband with increasing antennas, driving the need for efficient resource utilization\cite{b3}. Integrating communication and sensing into a unified platform allows for sharing resources like spectrum, energy, and hardware, improving efficiency and reducing costs. Thus, ISAC offers a promising solution for 6G, providing enhanced flexibility to support complex applications.
%第六代 (6G) 无线网络的发展有望支持智慧城市、自动驾驶和智能制造等新兴应用，具有大规模无线连接和高精度传感能力，可有效可靠地运行\cite{b1}。为了满足这些需求，集成传感和通信 (ISAC) 已成为 6G 网络的关键推动因素，引起了学术界和工业界的极大关注\cite{b2}。随着技术的进步，具有相似硬件和信号处理算法的无线通信和传感系统有望在高频宽带中运行，天线数量不断增加\cite{b3}。这将推动对更高效资源利用的需求。鉴于这些趋势，将通信和传感集成到统一平台中是切实可行的。它可以共享稀缺资源（例如频谱、能源和硬件），从而提高资源效率\cite{10666854}。此外，它降低了硬件成本和功耗，满足了对高性能系统和较低运营费用的需求。因此，ISAC 为未来的 6G 系统提供了一个有前景的解决方案，提供了增强的灵活性、可扩展性和效率，以支持 6G 的复杂应用。

Although ISAC has improved the synergy between communication and sensing, it still faces challenges, especially in dynamic environments and large-scale networks. For instance, {sensing capability of a single base station (BS)} is limited and vulnerable to interference and signal blockage in complex settings. To overcome these issues, the cooperative integrated communication and sensing (C-ISAC) network has emerged. By enabling collaboration among multiple {BSs} and sensor nodes, C-ISAC breaks the limitations of single-base-station systems, facilitating information sharing and cooperative sensing. This approach enhances target detection accuracy, optimizes resource allocation, and reduces interference and errors. Compared to traditional ISAC, C-ISAC offers greater communication robustness and more precise sensing in complex environments, significantly improving overall system performance.
%尽管 ISAC 提高了通信和传感之间的协同作用，但它仍然面临挑战，特别是在动态环境和大规模网络中。例如，单个基站的感知能力有限，并且在复杂环境中容易受到干扰和信号阻塞。为了克服这些问题，协作式集成通信和传感 (C-ISAC) 网络应运而生。通过实现多个基站和传感器节点之间的协作，C-ISAC 打破了单基站系统的限制，促进了信息共享和协作感知。这种方法提高了目标检测精度，优化了资源分配，并减少了干扰和错误。与传统 ISAC 相比，C-ISAC 在复杂环境中具有更高的通信稳健性和更精确的感知能力，从而显著提高了整体系统性能。

\textbf{Motivation and Challenges.} However, in C-ISAC networks, system design and optimization face several key challenges: (1) \textbf{Power consumption.} C-ISAC systems need to simultaneously support communication and sensing tasks, requiring the {BSs} to perform high-complexity signal processing and collaborative beamforming. In scenarios involving dynamic {movable} antenna adjustments and multi-target tracking, power consumption increases significantly, which is particularly critical for energy-constrained devices such as drones or IoT terminals. (2) \textbf{Time synchronization error.} In practical operations, clock misalignment between devices can introduce time synchronization (TS) errors. These errors degrade sensing accuracy, affect the stability of communication links, and have a significant adverse impact on overall system performance. (3) \textbf{CSI estimation error}. Due to the dynamic nature of wireless channels and the limited measurement capabilities, obtaining accurate channel state information (CSI) is highly challenging. CSI estimation errors lead to beamforming mismatches, reducing the signal-to-interference-plus-noise ratio (SINR) for communication and the resolution of sensing signals, thus impairing the performance of both communication and sensing in the integrated system.

To {address} these challenges, we propose a novel movable antenna (MA)-enabled C-ISAC system. The contributions of this paper are summarized as follows
\begin{itemize}
\item First, we reduce the feasible region of the communication rate constraint in the MA-aided C-ISAC system through a worst-case analysis and examine {an} impact of TS errors on this lower bound. We then use this lower bound to analyze the hybrid Cramér-Rao lower Bound (HCRLB) for target position estimation, considering the joint effects of TS errors and target position on accuracy. Our analysis shows that TS errors increase the HCRLB for target position estimation. Based on this, we propose a joint robust optimization strategy to minimize {transmit power} while ensuring the target position estimation meets the HCRLB requirements and satisfies the worst-case communication rate constraint. %首先，我们基于最坏情况分析缩小了 C-ISAC 系统中通信速率约束的可行域，并研究了 TS 误差对该下限的影响。随后，我们应用该下限来分析用于目标位置估计的混合 CRLB，突出了 TS 误差和目标位置对估计精度的联合影响。我们的分析表明，TS 误差会增加用于目标位置估计的 CRLB。基于这些发现，我们提出了一种联合优化策略，以最小化发射功率，同时确保目标位置估计满足混合 CRLB 要求并且通信速率满足最坏情况通信速率约束。
\item Next, we consider the problem of minimizing power while ensuring {target sensing accuracy} and {meeting} system rate constraints. Due to the highly non-convex nature of the problem, obtaining a global optimal solution is challenging. 

\item To address this non-convex problem, we first reformulate the objective function and constraints into a more tractable form using constraint deep reinforcement learning (CDRL) framework. Then, we use a modified deep deterministic policy gradient
(DDPG) algorithm with {the} Wolpertinger architecture for efficient
training under complex constraints. 
%接下来，我们考虑在确保 TS 精度和满足系统速率约束的同时最小化功率的问题。由于问题的高度非凸性，获得全局最优解具有挑战性。为了解决这个问题，我们首先使用约束深度强化学习 (CDRL) 将目标函数和约束重新表述为更易于处理的形式。然后，我们开发了一种基于惩罚的算法来有效地解决重新表述的问题并获得有效的解决方案。
\item Simulation results indicate that CSI and TS errors significantly impact both communication and sensing performance, thereby reducing the overall system efficiency. The results also demonstrate that the proposed MA-based approach outperforms traditional methods such as FPA under various configurations. This highlights the advantages of MA technology in terms of robustness and adaptability, confirming its effectiveness in mitigating CSI and TS errors while maintaining low power consumption. The proposed design not only saves $30\%$-$40\%$ {of the} power compared to existing algorithms but also shows {high} reliability in handling TS error variance of $100$~ns and CSI error of $0.01$.
%仿真结果表明，CSI 和 TS 误差对通信和传感性能都有重大影响，从而降低了系统的整体效率。结果还表明，所提出的基于移动天线的方法在各种设置下都优于 FPA 等传统方法。这凸显了移动天线技术在稳健性和适应性方面的优势，证实了其能够有效缓解 CSI 和 TS 误差，同时保持低功耗。所提出的设计不仅提高了系统性能，而且在处理这些关键错误时也表现出了增强的可靠性。
\end{itemize}

\textbf{Organization:} 
The paper is organized as follows: In \textbf{Section \ref{II}}, we present the system model for the MA-enabled C-ISAC system. \textbf{Section \ref{III}} derives the worst-case communication rate constraint for CSI errors and the HCRLB for TS errors, followed by the problem formulation. In \textbf{Section \ref{IV}}, we propose a CDRL algorithm to solve the problem and analyze its computational complexity. \textbf{Section \ref{VI}} provides simulation results that evaluate the performance of the proposed algorithm in the MA-enabled C-ISAC system. Finally, we conclude the paper in \textbf{Section \ref{V}}.

\section{LITERATURE REVIEW}

\subsection{ISAC and C-ISAC Communication Model}
The ISAC system integrates communication and sensing into a unified platform, optimizing resource utilization and enhancing performance. Transmit beamforming is crucial in these systems, as it directs signals toward the target, increasing spatial degrees of freedom (DoF) and improving both data transmission and sensing accuracy. Various studies have proposed beamforming strategies to enhance ISAC performance. For instance, \cite{b4} developed a joint beamforming scheme based on signal-to-interference-noise ratio (SINR) constraints to minimize sensing errors, while \cite{b5} focused on optimizing beamforming to maximize sensing performance under SINR requirements. However, these studies mainly address {single-BS} ISAC systems, which face coverage limitations due to obstacles and signal propagation losses.
To overcome these challenges, C-ISAC systems, which enable multi-BS collaboration, have gained attention. C-ISAC enhances spatial diversity for communication and multi-viewpoint sensing, reducing inter-cell interference, boosting communication rates, and improving target detection accuracy and resolution. In \cite{b6}, coordinated beamforming was optimized to maximize detection probability within a sensing region while meeting SINR constraints. \cite{b7} proposed joint optimization of transmit and receive beamforming to maximize sensing SINR while adhering to communication SINR requirements. 

%ISAC 系统将通信和感知集成到一个统一的平台中，优化资源并提高性能。在 ISAC 系统中，发射波束成形对于将信号引导至所需目标、增加空间自由度 (DoF) 以及提高数据传输和目标感知精度至关重要。许多研究提出了各种波束成形策略来增强 ISAC 性能。例如，在 \cite{b4} 中，设计了一种基于信号与干扰噪声比 (SINR) 约束的联合波束成形方案，以最大限度地减少感知误差。在 \cite{b5} 中，研究了波束成形优化以在满足 SINR 要求的同时最大限度地提高感知性能。然而，这些研究主要集中在由于障碍物和信号传播损耗而覆盖范围有限的单基站 (BS) ISAC 系统。为了克服这个问题，网络化 ISAC 系统 (C-ISAC) 引起了人们的关注。通过多基站协作，C-ISAC 增强了通信的空间分集并改善了多视点感知。多基站协作可以减少小区间干扰、提高通信速率并从不同观测角度提供更多的感知信息，从而提高目标检测精度和分辨率。在\cite{b6}中，优化了协调波束成形以在满足 SINR 约束的条件下最大化感知区域中的检测概率。在\cite{b7}中，提出了发射和接收波束成形的联合优化，以在通信 SINR 约束下最大化感知 SINR。在\cite{9729746}中，探索了基站选择、用户关联和波束成形的联合优化，以在保证感知性能的同时最小化发射功率。然而，这些研究大多假设理想条件，例如完美的信道估计和基站间的时间同步，而这在实际应用中很难实现。

In the C-ISAC system, both CSI estimation and TS errors can significantly degrade communication and sensing performance \cite{b8}. In C-ISAC systems with multi-BS collaboration \cite{b9}, CSI errors reduce beamforming gains and hinder resource allocation and coordination. From a sensing standpoint, multi-static sensing, which can be deployed without major network changes, is especially susceptible to TS errors, leading to performance degradation \cite{b10}. TS between BS clocks can cause significant deviations in sensing delay and distance estimation, resulting in ambiguities \cite{b11}. Sensing tasks demand much higher synchronization accuracy than communication; even nanosecond-level TS errors can lead to meter-level inaccuracies in distance measurements, which is critical for applications like autonomous driving and security monitoring. Therefore, accurate CSI estimation and precise TS are crucial to enhancing C-ISAC system performance.
%在 C-ISAC 系统中，信道状态信息 (CSI) 估计和时间同步 (TS) 误差会显著影响通信和感知性能\cite{b8}。在实际系统中，有限的训练资源和复杂的信道条件使准确的 CSI 估计变得困难，直接影响通信的可靠性和效率\cite{9505311}。在具有多基站协作的网络化 ISAC系统中\cite{b9}，CSI 误差会降低波束成形增益并损害资源分配和网络协作。从感知角度来看，多静态感知可以在不对网络进行重大更改的情况下部署，但它会引入显著的 TS 误差，从而降低系统性能\cite{b10}。基站时钟之间的时间不同步会导致感知延迟测量和距离估计出现较大偏差，从而可能导致模糊性\cite{b11}。此外，感知任务比通信需要更高的同步精度；例如在目标检测和距离估计中，即使是纳秒级的TS误差也会导致米级的距离估计误差，这对于自动驾驶、安防监控等高精度应用来说尤其成问题\cite{b33}。因此，保证高精度的CSI估计和TS对于提升C-ISAC系统的性能至关重要。

\subsection{C-ISAC network with TS error and CSI error}
Numerous studies have proposed optimization strategies to enhance the robustness of C-ISAC systems against CSI imperfections and TS issues \cite{b12,b13,b14}. For example, \cite{b12} introduced a resource allocation framework using variable-length snapshots to address CSI inaccuracies, while \cite{b13} proposed robust beamforming techniques to mitigate performance degradation due to inaccurate CSI.  Additionally, \cite{b14} optimized performance through coordinated transmit beamforming under synchronization constraints. Despite these advancements, challenges remain in large-scale, dynamic environments. Robust optimization techniques often involve high computational overhead, affecting real-time performance, while many synchronization strategies assume minimal delays, which can have significant impacts in large networks. Furthermore, no study has yet simultaneously addressed both CSI and synchronization errors to minimize power consumption in C-ISAC systems.
%许多研究提出了优化策略来解决 ISAC 系统的 CSI 稳健性和 TS 问题\cite{b12,b13,10659011,10534809,10649809,b14}。在\cite{b12} 中，作者引入了一种可变长度快照的资源分配框架来处理 CSI 缺陷。在\cite{b13} 中，提出了稳健的波束成形技术来减轻由于 CSI 不准确而导致的性能下降。在\cite{10659011} 中，作者提出了双稳健波束成形设计来考虑 CSI 和位置误差，从而提高动态环境中的系统可靠性。在\cite{10534809} 中，作者重点研究了水下物联网网络中的定位挑战，解决了异步、移动性和分层导致的同步问题。在 \cite{10649809} 中，引入了一种多层同步策略来抑制杂波并提高异步车载网络中的同步精度。在 \cite{b14} 中，提出了协调发射波束成形来优化同步条件下的性能。尽管取得了这些进展，但现有方法在大规模和高度动态的环境中仍然面临挑战。鲁棒优化技术通常会产生高计算开销，影响实际应用中的实时性能。许多同步策略假设时间延迟较小或偏移量最小，但在大规模网络中，异步误差和延迟会显著影响系统性能。此外，据我们所知，目前还没有一项研究同时解决 CSI 和同步误差以最大限度地降低 C-ISAC 系统的功耗。

\subsection{MA-enabled ISAC systems}
%许多研究都探索了支持 MA 的 ISAC 系统\cite{b15,b16,b17}。在\cite{b15} 中，作者考虑了具有 MA 的 ISAC 系统中的传统预编码，引入了一种稀疏优化方法来优化天线位置和预编码矩阵，以最大限度地减少用户间干扰和传输功率。在\cite{b16} 中，研究了 MA 增强型 ISAC 系统，通过优化天线位置来提高传输速率。在\cite{b17} 中，引入了一种新型 MA 辅助 ISAC 通信系统，优化了 MA 位置以提高通信容量和感知精度。与传统的 FPA MIMO 系统不同，该系统可以灵活地调整发射/接收 MA 的位置以重新配置 MIMO 信道，从而实现更高的容量。除了上述文献外，MA 技术已广泛应用于各种通信场景\cite{b18, b19, b20, li2024multi, b21}。尽管这些研究已经研究了 ISAC 和 MA，但据我们所知，具有 CSI 和 TS 的 MA 启用 C-ISAC 系统中的错误尚未被探索。为了解决这一空白，我们提出了一种新颖的 MA 启用 C-ISAC 系统。本文的贡献总结如下

Many studies have explored MA-enabled ISAC systems \cite{b15,b16,b17}. For instance, \cite{b15} introduced a sparse optimization method to optimize antenna positions and the precoding matrix in traditional ISAC systems with MA, aiming to minimize inter-user interference and {transmit power}. Similarly, \cite{b16} focused on improving transmission rates by optimizing antenna positions in MA-enhanced ISAC systems. In \cite{b17}, a novel MA-aided ISAC communication system was proposed, where the MA position was optimized to enhance both communication capacity and sensing accuracy. Unlike traditional fixed-position antenna (FPA) multiple-input multiple-output (MIMO) systems, this approach allows flexible adjustment of transmit/receive MAs, enabling the reconfiguration of MIMO channels for higher capacity. Beyond ISAC, MA technology has been widely applied in various communication scenarios \cite{b18, b19, b20, b21}. 

\subsection{Comparison between Our Work and Related Studies}
Despite extensive research on C-ISAC networks, TS and CSI errors, and the use of MA technology, no prior work has explored using MA to enhance C-ISAC performance while addressing TS and CSI errors. Integrating MA into C-ISAC systems can help mitigate these errors and improve robustness and reliability, offering a promising research direction. Our work highlights the unique contributions of MA technology: it reduces synchronization discrepancies between BSs by improving transmission and reception time alignment, enhancing both communication and sensing accuracy. Additionally, MA adapts to dynamic channel conditions by adjusting antenna positions based on real-time feedback, improving CSI estimation. This adaptability helps mitigate channel variations and timing errors, ensuring robust performance in C-ISAC networks.
%为了减少 TS 和 CSI 错误的影响，移动天线 (MA) 技术提供了一种有效的解决方案。MA 技术通过动态调整天线位置来实时优化信道环境。在 ISAC 系统中，移动天线通过改善发送和接收时间的对齐来减少基站之间的同步差异，从而提高通信和感知精度。此外，MA 技术通过根据实时反馈调整天线位置来适应不断变化的信道条件，从而改善 CSI 估计。这种适应性有助于减轻信道变化和定时误差的影响，确保在实际场景中实现稳健的性能。总体而言，ISAC 系统中的 MA 技术有效地解决了 TS 和 CSI 错误。

\section{System Model and Problem Formulation}\label{II}
\begin{figure}[htbp]
  \centering
  \includegraphics[scale=0.3]{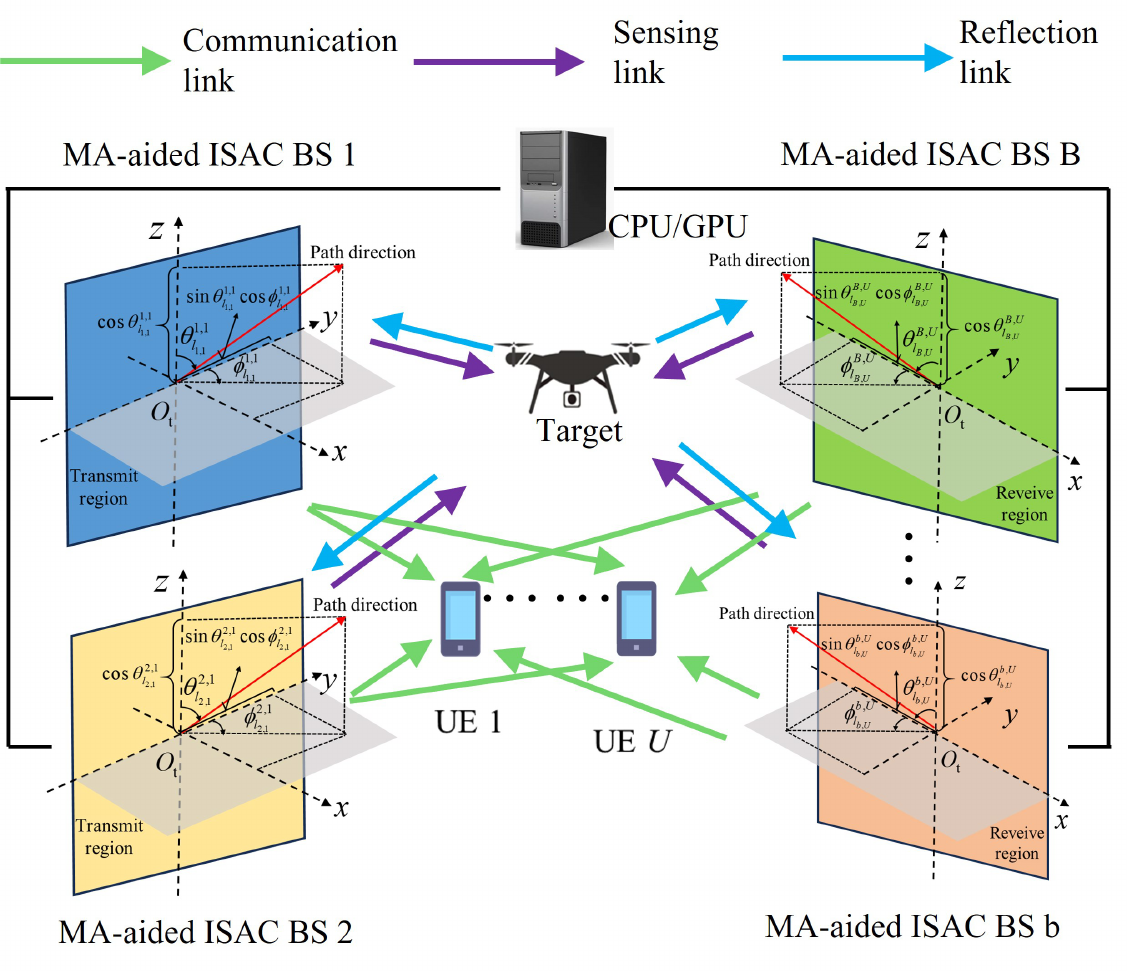}
  \captionsetup{justification=centering}
  \caption{Illustration of the MA-enabled C-ISAC system.}\vspace{-10pt}
\label{FIGURE0}
\end{figure}
\textbf{Transmit signal.} As shown in \textbf{Fig}.\ref{FIGURE0}, we consider a C-ISAC system that consists of $B$ dual-function radar and communication (DFRC) BSs. The set of BSs is denoted as $\mathcal{B}=\{1,\cdots,B\}$. Each BS is equipped with a uniform planar array (UPA) with $N=N_{x}\times N_{y}$ transmit MAs and $M=M_{x}\times M_{y}$ sensing receive MAs, where $N_{x}(M_{x})$ represents the number of horizontally transmit (receive sensing) MAs and the other $N_{y}(M_{y})$ represents the number of vertically transmit (receive sensing) MAs. The transmit set of MAs is represented as $\mathcal{N}=\{1,\cdots,N\}$, while the sensing receive set is defined as $\mathcal{M}=\{1,\cdots,M\}$. Each BS aims to transmit downlink signals to its associated users in the C-ISAC system. The BS collects echo signals reflected by point targets at its sensing receiver. Subsequently, all BSs forward this data to a central data processing center, where it undergoes joint target location estimation through data-level fusion. Common data-level approaches include maximum likelihood estimation (MLE) based fusion and weighted fusion techniques\cite{b22}. The integration of sensing information is crucial for multi-BS cooperative sensing\cite{b22}.
Integrating sensing information is crucial for multi-BS cooperative sensing. In the C-ISAC system, the coordinates of the $b$-th BS are represented as $\boldsymbol{p}=[x_{b},y_{b},z_{b}]^{T}$.
The C-ISAC system consists of $U$ single-antenna users and one sensing point target, where the set of users is denoted as $\mathcal{U}=\{1,\cdots,U\}$. We assume that the transmit signal of the $b$-th BS transmitter at time $n$ is given by
%如图 ref{FIGURE0} 所示，我们考虑一个由 $B$ 个双功能雷达和通信 (DFRC) 基站 (BS) 组成的 C-ISAC 系统。BS 集表示为 $\mathcal{B}=\{1,\cdots,B\}$。每个 BS 配备一个均匀平面阵列 (UPA)，其中 $N=N_{x}\times N_{y}$ 个发射可移动天线 (MA) 和 $M$ 个感测接收 MA，其中 $N_{x}$ 表示水平 MA 的数量，另一个 $N_{y}$ 表示垂直 MA 的数量。发射 MA 集表示为 $\mathcal{N}=\{1,\cdots,N\}$，而感测接收集定义为 $\mathcal{M}=\{1,\cdots,M\}$。在 C-ISAC 系统中，每个 BS 都旨在向其相关用户发送下行链路信号。BS 在其传感接收器上收集点目标反射的回波信号。随后，所有 BS 将此数据转发到中央数据处理中心，在那里通过数据级融合进行联合目标位置估计。常见的数据级方法包括基于最大似然估计 (MLE) 的融合和加权融合技术\cite{b22}。传感信息的集成对于多 BS 协作传感至关重要\cite{9551293}。集成传感信息对于多 BS 协作传感至关重要。在 C-ISAC 系统中，第 $b$ 个 BS 的坐标表示为 $\boldsymbol{p}=[x_{b},y_{b},z_{b}]^{T}$。C-ISAC 系统由 $U$ 个单天线用户和一个传感点目标组成，其中用户集表示为 $\mathcal{U}=\{1,\cdots,U\}$。 {为了简化演示并揭示多基站系统面临的小区间干扰的基本见解，考虑每个基站 ISAC 发射机为一个用户服务的设置，因此，我们设置 $M=U$。}我们假设第 $b$ 个基站发射机在时间 $n$ 的发射信号由下式给出
\begin{small}
\begin{align}
\boldsymbol{s}_{b}(n)=\boldsymbol{W}_{b}\boldsymbol{x}(n), b\in\mathcal{B},\label{pro1}
\end{align}    
\end{small}%
where $\boldsymbol{W}_{b}=[\boldsymbol{w}_{b,1},\cdots,\boldsymbol{w}_{b,U}]\in\mathbb{C}^{N\times U}$ and $\boldsymbol{x}(n)=[x_{1}(n),\cdots,x_{U}(n)]^{T}\in\mathbb{C}^{U\times 1}$ denotes the communication signal from the $b$-th BS, which follows an independent normal distribution with mean $0$ and variance $1$.
$\boldsymbol{W}_{b}$ represents the BS's beamforming. $n\in(0,S]$ and $S$ denotes the duration of the interval.

\textbf{MA channel.} Let $\boldsymbol{h}_{b,u}(\tilde{\boldsymbol{t}}_{b})\in\mathbb{C}^{N\times1}$ and $\boldsymbol{H}_{b^{\prime}, b}(\tilde{\boldsymbol{r}}_{b^{\prime}}, \tilde{\boldsymbol{t}}_{b})\in\mathbb{C}^{N\times M}$ represent the downlink communication channel and the sensing channel, respectively, where $\tilde{\boldsymbol{t}}_{b}=[\boldsymbol{t}_{b}^{1},\cdots,\boldsymbol{t}_{b}^{N}]\in\mathbb{R}^{3\times N}$, $\boldsymbol{t}_{b}^{n}=[x_{t,b}^{n},y_{t,b}^{n}]^{T}\in\mathcal{C}_{t,b}$ and $\tilde{\boldsymbol{r}}_{b}=[\boldsymbol{r}_{b}^{1},\cdots,\boldsymbol{r}_{b}^{M}]\in\mathbb{R}^{3\times N}$, $\boldsymbol{r}_{b}^{m}=[x_{r,b}^{m},y_{r,b}^{m}]\in\mathcal{C}_{r,b}$ denote the 3D coordinates of the positions of the communication transmit MA and the sensing receive MA of the $b$-th BS.
$\mathcal{C}_{t,b}=[x_{t,b}^{\min},x_{t,b}^{\max}]\times [y_{t,b}^{\min},y_{t,b}^{\max}]$ and $\mathcal{C}_{r,b}=[x_{r,b}^{\min},x_{r,b}^{\max}]\times [y_{r,b}^{\min},y_{r,b}^{\max}]$. Both 
$\boldsymbol{h}_{b,u}(\tilde{\boldsymbol{t}}_{b})$ and $\boldsymbol{H}_{b^{\prime}, b}(\tilde{\boldsymbol{r}}_{b^{\prime}}, \tilde{\boldsymbol{t}}_{b})$ are assumed to be quasi-static block fading channels, with multipath components concentrated within a given region at a specific fading block located at an arbitrary position\cite{b23}. 
$\boldsymbol{h}_{b,u}(\tilde{\boldsymbol{t}}_{b})$ and $\boldsymbol{H}_{b^{\prime}, b}(\tilde{\boldsymbol{r}}_{b^{\prime}}, \tilde{\boldsymbol{t}}_{b})$ are characterized by the elevation angle and azimuth angle of departure (AoD), angle of arrival (AoAs) and MA position. Based on the field response channel model \cite{b24}, 
$\boldsymbol{h}_{b,u}(\tilde{\boldsymbol{t}}_{b})$ and $\boldsymbol{H}_{b^{\prime}, b}(\tilde{\boldsymbol{r}}_{b^{\prime}}, \tilde{\boldsymbol{t}}_{b})$ are expressed as
\begin{small}
\begin{align}
&\boldsymbol{h}_{b,u}(\tilde{\boldsymbol{t}}_{b})=\tilde{\boldsymbol{h}}_{b,u}^{H}\boldsymbol{A}_{b,u}(\tilde{\boldsymbol{t}}_{b}),\nonumber\\
&\boldsymbol{H}_{b^{\prime}, b}(\tilde{\boldsymbol{r}}_{b^{\prime}}, \tilde{\boldsymbol{t}}_{b})=\alpha_{b}\boldsymbol{a}(\tilde{\boldsymbol{r}}_{b^{\prime}})^{H}\boldsymbol{a}(\tilde{\boldsymbol{t}}_{b}), b,b^{\prime}\in\mathcal{B}.\label{pro2}
\end{align}    
\end{small}%
Since the users are single-antenna, the field response matrix (FRM) only exists at the transmitting end of the communication channel. $\tilde{\boldsymbol{h}}_{b,u}\in\mathbb{C}^{L_{b,u}\times 1}$ is the complex channel gain, and $L_{b,u}$ represents the number of communication paths from the downlink the $b$-th BS to the $u$-th user. $\boldsymbol{A}_{b,u}(\tilde{\boldsymbol{t}}_{b})$ is the transmit FRM of the downlink channel. For the signal propagation distance difference of the 
$l_{b,u}$-th transmit path ($1\leq l_{b,u}\leq L_{b,u}$) between position $\tilde{\boldsymbol{t}}_{b}$ and the origin of the transmit region (i.e., $O_{b}$ in \textbf{Fig}.\ref{FIGURE0}), it can be expressed as $\rho_{l_{b,u}}^{b}(\boldsymbol{t}_{b}^{n})=x_{t,b}^{n}\cos\theta_{l_{b,u}}^{b}\cos\phi_{l_{b,u}}^{b}+y_{t,b}^{n}\cos\theta_{l_{b,u}}^{b}\sin\phi_{l_{b,u}}^{b}$, where the elevation and azimuth AoDs as $\theta_{l_{b,u}}^{b,u},\phi_{l_{b,u}}^{b,u}\in[0,\pi]$ respectively. 
Similarly, for the sensing signal between the target and the origin of the transmit region (i.e., $O_{b}$ in \textbf{Fig}.\ref{FIGURE0}), it can be expressed as $\rho^{b^{\prime}}(\boldsymbol{r}_{b^{\prime}}^{m})=x_{r,b}^{m}\cos\bar{\theta}^{b}\cos\bar{\phi}^{b}+y_{r,b}^{m}\cos\bar{\theta}^{b}\sin\bar{\phi}^{b}$, where the elevation and azimuth AoAs {are denoted as} $\bar{\theta}^{b},\bar{\phi}^{b}\in[0,\pi]$ respectively. According to the BS coordinates and target coordinates,  $\rho^{b}(\boldsymbol{t}_{b}^{n})$ and $\rho^{b^{\prime}}(\boldsymbol{r}_{b^{\prime}}^{m})$ can be rewritten as (\ref{pro9}) at the top of next page. Let $\lambda$ denote the carrier wavelength, and the phase difference of transmit MA and sensing receive MA can be calculated as $2\pi/\lambda\rho_{l_{b,u}}^{b}(\tilde{\boldsymbol{t}}_{b})$ and $2\pi/\lambda\rho^{b}(\tilde{\boldsymbol{r}}_{b})$, respectively. Therefore, the transmit and receive FRV characterizing the phase difference of $L_{b,u}$ transmit paths and sensing receive paths are expressed as
\begin{small}
\begin{align}
&\boldsymbol{a}_{b,u}(\boldsymbol{t}_{b}^{n})=\left[e^{j2\pi/\lambda\rho_{1}^{b}(\boldsymbol{t}_{b}^{n})},\cdots,e^{j2\pi/\lambda\rho_{L_{b,u}}^{b}(\boldsymbol{t}_{b}^{n})}\right]^{T}\in\mathbb{C}^{L_{b,u}\times 1},\nonumber\\
&a(\boldsymbol{r}_{b}^{m})=e^{j2\pi/\lambda\rho^{b^{\prime}}(\boldsymbol{r}_{b^{\prime}}^{m})}\in\mathbb{C}^{1\times 1}.\label{pro3}
\end{align}    
\end{small}
Then, $\boldsymbol{A}_{b,u}(\tilde{\boldsymbol{t}}_{b})$ and $\boldsymbol{a}(\tilde{\boldsymbol{r}}_{b})$ are computed as
\begin{small}
\begin{align}
&\boldsymbol{A}_{b,u}(\tilde{\boldsymbol{t}}_{b})=[\boldsymbol{a}_{b,u}(\boldsymbol{t}_{b}^{1}),\cdots,\boldsymbol{a}_{b,u}(\boldsymbol{t}_{b}^{N})]\in\mathbb{C}^{L_{b,u}\times N}\nonumber\\
&\boldsymbol{a}(\tilde{\boldsymbol{r}}_{b})=\left[e^{j2\pi/\lambda\rho_{1}^{b}(\boldsymbol{r}_{b}^{1})},\cdots,e^{j2\pi/\lambda\rho^{b^{\prime}}(\boldsymbol{r}_{b^{\prime}}^{m})}\right]^{T}\in\mathbb{C}^{M\times 1}.\label{pro4}
\end{align}    
\end{small}%
Similarly, the transmit FRV for the sensing channel is expressed as
\begin{small}
\begin{align}
\boldsymbol{a}(\tilde{\boldsymbol{t}}_{b})=\left[e^{j2\pi/\lambda\rho_{1}^{b}(\boldsymbol{t}_{b}^{1})},\cdots,e^{j2\pi/\lambda\rho^{b}(\boldsymbol{t}_{b}^{N})}\right]^{T}\in\mathbb{C}^{M\times 1}.\label{pro5}
\end{align}    
\end{small}%
%where 
%$\bar{\rho}^{b}(\boldsymbol{t}_{n})=x_{n}^{b}\sin\theta^{b}\cos\phi^{b}+y_{n}^{b}\cos\theta^{b}$ represent the signal propagation distance differences between the position of the sensing transmit path and the origin of the transmit region (i.e., $O_{b}$ in Fig. 2). and 
%$\bar{\rho}^{b}(\boldsymbol{r}_{m})=x_{m}^{b}\sin\theta^{b}\cos\phi^{b}+y_{m}^{b}\cos\theta^{b}$ represent the signal propagation distance differences between position 
%at the sensing transmit path, sensing receive path, sensing and the origin of the receive region $O_{r}$.  
Moreover, $\alpha_{b}$ denotes the reflection coefficient incorporating the effects of the radar cross section (RCS)  from the transmit MAs at the $b$-th BS to the target to the sensing receive MAs at the $b^{\prime}$-th BS. This study investigates multi-static sensing in a C-ISAC system, where the BS ISAC transmitter and receiver are co-located. The receiver collects target reflection signals, and sends the results to the cloud server for collaborative sensing through data fusion. We focus on the stage where the BS has prior knowledge of the target parameters, obtained during detection. Using these parameters, the BS ISAC transmitter optimizes beamforming for target estimation and other sensing tasks. The cloud server also helps the BS determine its location, while the receiver reduces unwanted signals from the LoS path and clutter from static objects, assuming known NLoS paths. An LoS channel model in (\ref{pro2}) is used to describe the BS-target link.

\subsection{Cooperative Communication Model}
In the $n$-th time slot, the receive signal for the $u$-th user is represented as
\begin{small}
\begin{align}
&y_{u}(n)=\sum\nolimits_{b=1}^{B}c_{b,u}\boldsymbol{h}_{b,u}(\tilde{\boldsymbol{t}}_{b})\boldsymbol{s}_{b}(n)+z_{u}(n),~c_{b,u}\in\{0,1\},\label{pro6}
\end{align}
\end{small}%
where $z_{u}(n)\sim\mathcal{CN}(0,\sigma_{u}^{2})$ represents the additive white Gaussian noise (AWGN) experienced by user $u$. Equation (\ref{pro6}) shows that each user is influenced by interference from neighbouring BSs, highlighting the need for coordinated transmission beamforming to reduce this interference. In this context, $c_{b,u}$ serves as the selection variable. This {communication link} is not established if $c_{b,u}=0$ indicates the downlink from the $b$-th BS to the $u$-th user. In contrast, this {communication link} is established if $c_{b,u}=1$ represents the downlink from the $b$-th BS to the $u$-th user.
%其中 $z_{u}(n)\sim\mathcal{CN}(0,\sigma_{u}^{2})$ 表示用户 $u$ 所经历的加性高斯白噪声 (AWGN)。方程 (\ref{pro6}) 表明每个用户都会受到邻近基站 (BS) 干扰的影响，这凸显了协调传输波束成形以减少这种干扰的必要性。在这种情况下，$c_{b,u}$ 用作选择变量。如果 $c_{b,u}=0$ 表示从第 $b$ 个 BS 到第 $u$ 个用户的下行链路，则不建立此连接。相反，如果 $c_{b,u}=1$ 表示从第 $b$ 个 BS 到第 $u$ 个用户的下行链路，则建立此连接。

\subsection{Cooperative Sensing Model}
%BS 传感接收器分析接收到的目标反射信号，并将估计结果传输到中央处理单元 (CPU)。
%利用先前的目标参数和波束成形设计增强目标估计，同时考虑 LoS 信道模型而不考虑杂波和非目标信号\cite{10141975}。
%因此，第 $b$ 个基站在第 $n$ 个时隙期间接收到的传感信号表示为
The BS sensing receiver analyzes the receive target reflection signals and transmits the estimated results to the central processing unit (CPU).
Utilizing previous target parameters and beamforming design enhances target estimation while considering the LoS channel model without considering clutter and non-target signals\cite{b25}. 
Thus, the receive sensing signal at the $b$-th BS during the $n$-th time slot is denoted as
\begin{small}
\begin{align}
&\boldsymbol{y}_{b}(n)=\sum\nolimits_{b^{\prime}=1}^{B}\tilde{c}_{b}\boldsymbol{H}_{b^{\prime}, b}(\tilde{\boldsymbol{r}}_{b^{\prime}},\tilde{\boldsymbol{t}}_{b})\boldsymbol{s}_{b}(n-\tau_{b,b^{\prime}})+\boldsymbol{z}_{b}(n),\nonumber\\
&~\tau_{b,b^{\prime}}=\left\{\begin{matrix}
0,&b=b^{\prime}\\
\tau_{b,b^{\prime}},&b\neq b^{\prime}
\end{matrix}\right.~b,b^{\prime}\in\mathcal{B}, \label{pro7}
\end{align}
\end{small}%
where $t_{b^{\prime},b}$ denotes the propagation delay measurement from the 
$b$-th BS ISAC transmitter to the target and from the target to the 
$b^{\prime}$-th BS sensing receiver\cite{b25}, and $\tau_{b,b^{\prime}}$ is given by
%其中 $t_{b^{\prime},b}$ 表示从第 $b$ 个 BS ISAC 发射机到目标以及从目标到第 $b^{\prime}$ 个 BS 传感接收机的传播延迟测量\cite{b25}，而 $\tau_{b,b^{\prime}}$ 由下式给出
\begin{small}
\begin{align}
\tau_{b,b^{\prime}}=\hat{t}_{b,b^{\prime}}+\xi_{b,b^{\prime}}+\bar{\xi}_{b,b^{\prime}}, \xi_{b,b^{\prime}}=\Delta\xi_{b}-\Delta\xi_{b^{\prime}},\label{pro8}
\end{align}
\end{small}%
where $\Delta\xi_{b}$ and $\Delta\xi_{b^{\prime}}$ denote the TS error between the $b$/$b^{\prime}$-th BS and the reference clock. According to \cite{b25}, the TS error $\xi_{b,b^{\prime}}$ is a random variable, and it can be modeled as following a Gaussian distribution with mean $0$ and variance $\sigma^{2}_{\xi}$. $\hat{t}_{b,b^{\prime}}$ represents the transmission delay from the $b$-th BS, passing through the target and reflecting the $b^{\prime}$-th BS. $\bar{\xi}_{b,b^{\prime}}$ is the measurement noise. Finally, all mathematical symbols are defined in \textbf{TABLE}~\ref{tab:notation}.
%其中 $\Delta\xi_{b}$ 和 $\Delta\xi_{b^{\prime}}$ 表示第 $b$/$b^{\prime}$ 个基站与参考时钟之间的 TS 误差。根据 \cite{b25}，TS 误差 $\xi_{b,b^{\prime}}$ 是一个随机变量，可以将其建模为服从均值为 $0$、方差为 $\sigma^{2}_{\xi}$ 的高斯分布。$\hat{t}_{b,b^{\prime}}$ 表示从第 $b$ 个基站经过目标并反射到第 $b^{\prime}$ 个基站的传输延迟。$\bar{\xi}_{b,b^{\prime}}$ 是测量噪声。最后，所有数学符号均在 \textbf{TABLE}~\ref{tab:notation} 中定义。
\begin{figure*}
\begin{small}
\begin{align}
&\rho^{b}(\boldsymbol{t}_{b}^{n})=x_{n}^{b}\sin\underbrace{\left(\arctan\left((x_{T}-x_{b})/(x_{T}-y_{b})\right)+\pi\right)}_{\theta_{b}}\cos\underbrace{\left(\arctan\left(z_{b}/\sqrt{(x_{T}-x_{b})^{2}+(x_{T}-y_{b})^{2}}\right)+\pi\right)}_{\phi_{b}}+y_{n}^{b}\cos(\arctan((x_{T}-x_{b})/\nonumber\\
&(x_{T}-y_{b}))+\pi)\sin\left(\arctan\left(z_{b}/\sqrt{(x_{T}-x_{b})^{2}+(x_{T}-y_{b})^{2}}\right)+\pi\right), \nonumber\\
&\rho^{b^{\prime}}(\boldsymbol{r}_{b^{\prime}}^{m})=x_{m}^{b^{\prime}}\sin\underbrace{\left(\arctan\left((x_{T}-x_{b^{\prime}})/(x_{T}-y_{b^{\prime}})\right)+\pi\right)}_{\bar{\theta}_{b^{\prime}}}\cos\underbrace{\left(\arctan\left(z_{b^{\prime}}/\sqrt{(x_{T}-x_{b^{\prime}})^{2}+(x_{T}-y_{b^{\prime}})^{2}}\right)+\pi\right)}_{\bar{\phi}_{b}}+y_{m}^{b^{\prime}}\cos(\arctan((x_{T}-x_{b^{\prime}})/\nonumber\\
&(x_{T}-y_{b^{\prime}}))+\pi)\sin\left(\arctan\left(z_{b^{\prime}}/\sqrt{(x_{T}-x_{b^{\prime}})^{2}+(x_{T}-y_{b^{\prime}})^{2}}\right)+\pi\right).\label{pro9}
\end{align}     
\end{small}
\hrulefill
\end{figure*}

\begin{table}[h!]
\centering
\begin{tabular}{|l|l|}
\hline
\textbf{Symbol} & \textbf{Description} \\ \hline
$\mathcal{B}/\mathcal{U}$ & Set of all BSs/users \\ \hline
$B/U$ & Number of BSs/users \\ \hline
$\mathcal{N}/\mathcal{M}$ & Set of all transmit MAs/receive MAs \\ \hline
$N/M$ & Number of transmit MAs/receive MAs \\ \hline
$\boldsymbol{W}_{b}$ & Beamforming of the $b$-th BS \\ \hline
$S$ & Symbol duration  \\ \hline
$\tilde{\boldsymbol{t}}_{b}/\tilde{\boldsymbol{r}}_{b}$ & Position of transmit MAs/receive MAs \\ \hline
$\mathcal{C}_{t,b}/\mathcal{C}_{r,b}$ & Movable region of transmit MAs/receive\\
&MAs \\ \hline
$\boldsymbol{h}_{b,u}(\tilde{\boldsymbol{t}}_{b})$ & Channel BS-to-user\\ \hline
$\boldsymbol{H}_{b^{\prime}, b}(\tilde{\boldsymbol{r}}_{b^{\prime}}, \tilde{\boldsymbol{t}}_{b})$ & Channel BS-to-BS \\ \hline
$L_{b,u}$ & Number of channel paths \\ \hline
$\theta_{l_{b,u}}^{b,u}$/$\phi_{l_{b,u}}^{b,u}$ & Elevation and azimuth AoDs of\\
& communication signal \\ \hline
$\bar{\theta}^{b}$/$\bar{\phi}^{b}$ & Elevation and azimuth AoDs of sensing signal \\ \hline
$\theta^{b}$/$\phi^{b}$ & Elevation and azimuth AoAs of sensing signal \\ \hline
$c_{b,u}/\tilde{c}_{b}$ & Selection factor\\ \hline
$t_{b^{\prime},b}$ & Propagation delay measurement\\ \hline
$\Delta\xi_{b}$/$\Delta\xi_{b^{\prime}}$ & TS error\\\hline
\end{tabular}
\caption{Summary of {Notations}.}\vspace{-10pt}
\label{tab:notation}
\end{table}

\section{C-ISAC with Imperfect CSI and TS error}\label{III}
The practical deployment of the C-ISAC system depends {heavily} on the availability of CSI and TS between different BSs. However, due to imperfect CSI and TS data at these BSs, {both are prone to errors.} This section analyzes the worst-case communication rate constraints related to CSI estimation errors and the HCRLB concerning TS errors and their impact on sensing accuracy under imperfect conditions.
%C-ISAC 系统的实际部署在很大程度上取决于不同 BS 之间 CSI 和 TS 的可用性。然而，由于这些 BS 上的 CSI 和 TS 数据不完善，这两个元素都容易出错。本节分析了与 CSI 估计误差相关的最坏情况通信速率约束和与 TS 误差有关的克拉美-罗下限 (CRLB) 及其对不完善条件下感知精度的影响。
\subsection{Worst-case Achievable Rate with Imperfect CSI}
%根据\cite{b26}，引入通信速率约束对于确保系统满足所需的吞吐量和可靠性性能是必要的。通过纳入这一约束，我们可以确保通信系统在可用带宽和功率资源的限制范围内运行，同时为用户提供可接受的服务质量 (QoS)\cite{b27}。我们的目标是平衡频谱效率和系统稳健性之间的权衡，特别是在存在干扰和噪声的情况下。这种平衡对于优化资源分配和在实际通信系统中实现所需的性能水平至关重要\cite{b14}。总之，本文讨论了引入一种新的速率约束。根据接收信号表达式 (\ref{pro6})，通信速率约束表示为
According to \cite{b26}, introducing a communication rate constraint is essential to ensure the system meets required throughput and reliability. This constraint ensures the system operates within available bandwidth and power {while maintaining acceptable quality of service (QoS) for users \cite{b27}.} We aim to balance power consumption and system robustness, particularly in the presence of interference and noise. This balance is {the key} to optimizing resource allocation and achieving desired performance in practical systems \cite{b14}. In summary, this paper introduces a {communication} rate constraint, which is expressed as based on the receive signal in (\ref{pro6})
\begin{small}
\begin{align}
&R_{u}=\log_{2}\left(1+\sum\nolimits_{b=1}^{B}|c_{b,u}\boldsymbol{h}_{b,u}(\tilde{\boldsymbol{t}}_{b})\boldsymbol{w}_{b,u}|^{2}/\left(\sum\nolimits_{b=1}^{B}\sum\nolimits_{u^{\prime}=1,u^{\prime}\neq u}^{U}\right.\right.\nonumber\\
&\left.\left.|c_{b,u^{\prime}}\boldsymbol{h}_{b,u^{\prime}}(\tilde{\boldsymbol{t}}_{b})\boldsymbol{w}_{b,u^{\prime}}|^{2}\right)+\sigma_{b}^{2}\right)\geq\gamma_{u}.\label{pro10}
\end{align}   
\end{small}%   
%根据\cite{b28}，实际 CSI、估计 CSI 误差和真实 CSI 之间的关系表示为
Based on\cite{b28}, the relationship between the actual CSI, the estimated CSI error, and the true CSI is expressed as
\begin{small}
\begin{align}
\boldsymbol{h}_{b,u}(\tilde{\boldsymbol{t}}_{b})=\hat{\boldsymbol{h}}_{b,u}(\tilde{\boldsymbol{t}}_{b})+\Delta\boldsymbol{h}_{b,u}, \|\Delta\boldsymbol{h}_{b,u}\|\leq\epsilon_{b,u}.\label{pro11}
\end{align}
\end{small}%
%现有的基于MA的信道估计技术通过获取AoA、AoD和信道增益来重构信道。因此，我们有
The existing MA-based channel estimation techniques reconstruct the channel by obtaining the AoA, AoD, and channel gain\cite{b29}. Therefore, we have 
\begin{small}
\begin{align}
&\theta_{l_{b,u}}^{b}=\hat{\theta}_{l_{b,u}}^{b}+\Delta\theta_{l_{b,u}}^{b}, \phi_{l_{b,u}}^{b}=\hat{\phi}_{l_{b,u}}^{b}+\Delta\phi_{l_{b,u}}^{b},
\tilde{\boldsymbol{h}}_{b,u}=\hat{\tilde{\boldsymbol{h}}}_{b,u}+\Delta\tilde{\boldsymbol{h}}_{b,u},\nonumber\\
&|\Delta\theta_{l_{b,u}}^{b}|<\epsilon_{\theta}, |\Delta\phi_{l_{b,u}}^{b}|<\epsilon_{\phi}, \|\Delta\tilde{\boldsymbol{h}}_{b,u}\|\leq\bar{\epsilon}_{b,u}.\label{pro_11}
\end{align}
\end{small}%
Although $\epsilon_{\theta}$, $\epsilon_{\phi}$ and $\bar{\epsilon}_{b,u}$ are known, due to the nonlinear relationship between the estimation errors $\epsilon_{\theta}$, $\epsilon_{\phi}$ and $\bar{\epsilon}_{b,u}$ and true channel $\boldsymbol{h}_{b,u}(\tilde{\boldsymbol{t}}_{b})$, the range of the channel estimation error $\epsilon_{b,u}$ remains uncertain.
According to (\ref{pro11}), communication rate constraint (\ref{pro11}) is rewritten as
%虽然$\epsilon_{\theta}$、$\epsilon_{\phi}$和$\bar{\epsilon}_{b,u}$已知，但是由于估计误差$\epsilon_{\theta}$、$\epsilon_{\phi}$和$\bar{\epsilon}_{b,u}$与真实信道$\boldsymbol{h}_{b,u}(\tilde{\boldsymbol{t}}_{b})$之间存在非线性关系，因此信道估计误差$\epsilon_{b,u}$的范围仍然不确定。根据(\ref{pro11})，通信速率约束(\ref{pro11})可重写为
\begin{small}
\begin{align}
&R_{u}=\log_{2}\left(1+\sum\nolimits_{b=1}^{B}|c_{b,u}(\hat{\boldsymbol{h}}_{b,u}(\tilde{\boldsymbol{t}}_{b})+\Delta\boldsymbol{h}_{b,u})^{H}\boldsymbol{w}_{b,u}|^{2}/\left(\sum\nolimits_{b=1}^{B}\right.\right.\nonumber\\
&\left.\left.\sum\nolimits_{u^{\prime}=1,u^{\prime}\neq u}^{U}|c_{b,u^{\prime}}(\hat{\boldsymbol{h}}_{b,u^{\prime}}(\tilde{\boldsymbol{t}}_{b})+\Delta\boldsymbol{h}_{b,u^{\prime}})^{H}\boldsymbol{w}_{b,u^{\prime}}|^{2}\right)+\sigma_{b}^{2}\right)\geq\gamma_{u}.\label{pro12}   
\end{align}   
\end{small}%
%从 \cite{b28} 中的通信速率约束表达式中，我们注意到 $\Delta\boldsymbol{h}_{b,u^{\prime}}$ 出现在干扰项和信号增益项中。根据 \cite{b28} 中推导的最坏情况稳健理论，如果最坏情况通信速率约束的可行域小于实际通信速率约束的可行域，则优化过程中遇到的最坏情况通信速率约束将比实际最坏情况约束更具限制性。这种情况导致通信速率约束的可行域更加稳健。较小的可行集意味着降低了问题对输入数据的变化或不确定性的敏感度。通过更少的选项，数据波动对最优解的影响被最小化，从而导致更一致的性能\cite{b28}。因此，我们提出以下定理。
{From the communication rate constraint in \cite{b28},} we note that $\Delta\boldsymbol{h}_{b,u^{\prime}}$ appears in both the interference and signal gain terms. According to the worst-case {robust optimization} theory derived in\cite{b28}, if the feasible region of the worst-case communication rate constraint is smaller than that of the actual communication rate constraint, then the worst-case communication rate constraint encountered during the optimization process will be more restrictive than the actual worst-case constraint. This situation leads to a more robust and feasible region for the communication rate constraint. A smaller feasible set means which reduces the sensitivity of the problem to variations or uncertainties in the input data. With fewer options, the influence of data fluctuations on the optimal solution is minimized, resulting in more consistent performance\cite{b28}. Consequently, we present the following theorem.
\begin{theorem}\label{th1}
The communication rate constraint in (\ref{pro12}) can be scaled as a worst-case communication rate constraint, and it is expressed as
\begin{small}
\begin{align}
&\underbrace{A(\boldsymbol{w}_{b,u})/B(\boldsymbol{w}_{b,u^{\prime}})}_{C(\boldsymbol{w}_{b,u},\boldsymbol{w}_{b,u^{\prime}})}\geq(2^{\gamma_{u}}-1), \label{pro_13} 
\end{align}    
\end{small}%
{where} $A(\boldsymbol{w}_{b,u})$ and $B(\boldsymbol{w}_{b,u^{\prime}})$ are given by
\begin{small}
\begin{align}
&A(\boldsymbol{w}_{b,u})=\nonumber\\
&\sum\nolimits_{b=1}^{B}(c_{b,u}|\hat{\boldsymbol{h}}_{b,u}^{H}(\tilde{\boldsymbol{t}}_{b})\boldsymbol{w}_{b,u}|)^{2}-(c_{b,u}(N\|\hat{\tilde{\boldsymbol{h}}}_{b,u}\|^{2}+2NL_{b,u}\bar{\epsilon}_{b,u}))\nonumber\\
&|\boldsymbol{w}_{b,u}|^{2})-2c_{b,u}^{2}|\hat{\boldsymbol{h}}_{b,u}^{H}(\tilde{\boldsymbol{t}}_{b})\boldsymbol{w}_{b,u}|\sqrt{(N\|\hat{\tilde{\boldsymbol{h}}}_{b,u}\|^{2}+2NL_{b,u}\bar{\epsilon}_{b,u})\|\boldsymbol{w}_{b,u}\|^{2}}\nonumber\\
&B(\boldsymbol{w}_{b,u^{\prime}})=\left(\sum\nolimits_{b=1}^{B}\right.\sum\nolimits_{u^{\prime}=1,u^{\prime}\neq u}^{U}(c_{b,u^{\prime}}|\hat{\boldsymbol{h}}_{b,u^{\prime}}^{H}(\tilde{\boldsymbol{t}}_{b})\boldsymbol{w}_{b,u^{\prime}}|)^{2}+c_{b,u}^{2}\nonumber\\
&(N\|\hat{\tilde{\boldsymbol{h}}}_{b,u^{\prime}}\|^{2}+2NL_{b,u^{\prime}}\bar{\epsilon}_{b,u^{\prime}})\left.\left.\times\|\boldsymbol{w}_{b,u^{\prime}}\|^{2}+2c_{b,u^{\prime}}^{2}|\hat{\boldsymbol{h}}_{b,u^{\prime}}^{H}(\tilde{\boldsymbol{t}}_{b})\boldsymbol{w}_{b,u^{\prime}}|\right.\right.\nonumber\\
&\left.\left.\sqrt{(N\|\hat{\tilde{\boldsymbol{h}}}_{b,u^{\prime}}\|^{2}+2NL_{b,u^{\prime}}\bar{\epsilon}_{b,u^{\prime}})\|\boldsymbol{w}_{b,u^{\prime}}\|^{2}}\right)+\sigma_{b}^{2}\right). \label{pro13} 
\end{align}    
\end{small}%
This proof is given in \textbf{Appendix}~\textbf{A}. 
\end{theorem}

\subsection{Sensing Accuracy with TS Errors}
%在本部分中，我们重点关注目标位置 $\boldsymbol{p}=[p_{x}, p_{y}]^{T}$作为感兴趣的参数。CRLB 用于量化估计性能，作为任何无偏估计量的方差下限。根据 \textbf{Section~\ref{II}}，TS 误差是不可避免的，并且会显著影响目标估计。因此，我们推导出具有 TS 误差的 CRLB，以建立精确的性能界限。具体而言，每个 BS 接收器对接收信号$\bar{S}$ 在 $S$ 上进行采样，产生样本 ${\boldsymbol{y}_{b}(n_{1}),\cdots, \boldsymbol{y}_{b}(n_{\bar{S}})}$。让 $\mathcal{S}=\{n_{1},\cdots, n_{\bar{S}}\}$ 表示采样时间。频域信号$\boldsymbol{y}_{b}(n_{s})$通过离散傅里叶变换（DFT）\cite{b8}获得，$\boldsymbol{y}_{b}(n_{s})$的频域信号，$\forall t_{s}\in\mathcal{S}$可表示为
In {the following}, we focus on the target position $\boldsymbol{p}=[p_{x}, p_{y}]^{T}$
as the interested parameters. 
The Cramér-Rao Lower Bound (CRLB) is used to quantify estimation performance as the lower bound on the variance of any unbiased estimator. Based on \textbf{Section~\ref{II}}, TS errors are inevitable and significantly affect target estimation. Therefore, we derive the CRLB with TS errors to establish {tight} performance bounds. Specifically, each BS sensing receiver samples the receive signal
$\bar{S}$ over $S$, yielding samples ${\boldsymbol{y}_{b}(n_{1}),\cdots, \boldsymbol{y}_{b}(n_{\bar{S}})}$. Let $\mathcal{S}=\{n_{1},\cdots, n_{\bar{S}}\}$ denote the sampling times. The frequency domain signal 
$\boldsymbol{y}_{b}(n_{s})$ is obtained via discrete Fourier transform (DFT)\cite{b8}, the frequency-domain signal of $\boldsymbol{y}_{b}(n_{s})$, $\forall t_{s}\in\mathcal{S}$ is obtained as
\begin{small}
\begin{align}
\bar{\boldsymbol{y}}_{b}(f_{s})=\sum\nolimits_{b^{\prime}=1}^{B}\bar{c}_{b}\boldsymbol{H}_{b,b^{\prime}}(\tilde{\boldsymbol{r}}_{b^{\prime}},\tilde{\boldsymbol{t}}_{b})\boldsymbol{W}_{b}e^{-jf_{s}\tau_{b,b^{\prime}}}\boldsymbol{s}_{b}(f_{s})+\bar{\boldsymbol{z}}_{b}(f_{s}).\label{pro14}
\end{align}
\end{small}%
%然后，第 $b$ 个 ISAC BS 感知接收器接收到的信号在频域中表示为
Then, the signal receive by the $b$-th ISAC BS sensing receiver is expressed in the frequency domain as 
\begin{small}
\begin{align}
&\underbrace{\left[\begin{matrix}
\bar{\boldsymbol{y}}_{b}(f_{1})\\
\vdots\\
\bar{\boldsymbol{y}}_{b}(f_{\bar{S}})
\end{matrix}\right]}_{\tilde{\boldsymbol{y}}_{b}}
=\underbrace{\left[\begin{matrix}
\sum\nolimits_{b^{\prime}=1}^{B}\bar{c}_{b}\boldsymbol{H}_{b,b^{\prime}}(\tilde{\boldsymbol{r}}_{b^{\prime}},\tilde{\boldsymbol{t}}_{b})\boldsymbol{W}_{b}e^{-jf_{1}\tau_{b,b^{\prime}}}\boldsymbol{s}_{b}(f_{1})\\
\vdots\\
\sum\nolimits_{b^{\prime}=1}^{B}\bar{c}_{b}\boldsymbol{H}_{b,b^{\prime}}(\tilde{\boldsymbol{r}}_{b^{\prime}},\tilde{\boldsymbol{t}}_{b})\boldsymbol{W}_{b}e^{-jf_{\bar{S}}\tau_{b,b^{\prime}}}\boldsymbol{s}_{b}(f_{\bar{S}})\\
\end{matrix}\right]}_{\tilde{\boldsymbol{x}}_{b}}+\tilde{\boldsymbol{z}}_{b},\label{pro15}
\end{align}    
\end{small}%
%其中 $\tilde{\boldsymbol{z}}_{b}=[\bar{\boldsymbol{z}}_{b}(f_{1}),\cdots,\bar{\boldsymbol{z}}_{b}(f_{\bar{S}})]^{T}$，$\tilde{\boldsymbol{z}}_{b}\sim\mathcal{CN}(0,\sigma_{b}^{2}\boldsymbol{I})$。根据(\ref{pro15})中的频域感知信号表达式，位置参数
%$\boldsymbol{p}_{T}$和同步参数
%$\boldsymbol{\xi}_{b}=[\xi_{1,b},\cdots,\xi_{B,b}]^{T}\sim\mathcal{N}(\boldsymbol{0},\sigma_{\xi}^{2}\boldsymbol{I})$均为未知数，此时需要利用第$b$个基站的感知接收信号对共$B+2$个参数进行联合估计。众所周知，混合Cramér-Rao下界(HCRLB)更适合用于确定多源参数估计精度的下界。由于 HCRLB 为确定性参数的任何无偏估计量和随机参数的任何估计量提供了均方误差 (MSE) 的下限，因此推导参数 $\boldsymbol{p}_{T}$ 和 $\boldsymbol{\xi}_{b}$ 的 HCRLB 需要首先根据传感接收信号确定 MSE。因此，基于传感接收信号
%$\tilde{\boldsymbol{y}}_{b}$ 的位置参数
%$\boldsymbol{p}_{T}$ 和同步参数 $\boldsymbol{\xi}_{b}$ %的 MSE 下限表示如下
where $\tilde{\boldsymbol{z}}_{b}=[\bar{\boldsymbol{z}}_{b}(f_{1}),\cdots,\bar{\boldsymbol{z}}_{b}(f_{\bar{S}})]^{T}$, $\tilde{\boldsymbol{z}}_{b}\sim\mathcal{CN}(0,\sigma_{b}^{2}\boldsymbol{I})$. According to the frequency-domain sensing signal expression in (\ref{pro15}), both the position parameter 
$\boldsymbol{p}_{T}$ and synchronization parameter 
$\boldsymbol{\xi}_{b}=[\xi_{1,b},\cdots,\xi_{B,b}]^{T}\sim\mathcal{N}(\boldsymbol{0},\sigma_{\xi}^{2}\boldsymbol{I})$ are unknown. In this case, the sensing receiver signal at the $b$-th BS is used to jointly estimate a total of $B+2$ parameters. As is well known, the HCRLB is more suitable for determining the lower bound of the estimation accuracy for multi-source parameters{\cite{b30,b31}}. Since the HCRLB provides a lower bound on the mean squared error (MSE) for any unbiased estimator of deterministic parameters and any estimator of stochastic parameters, deriving the HCRLB for parameters $\boldsymbol{p}_{T}$ and $\boldsymbol{\xi}_{b}$ requires first determining the MSE based on the sensing receive signal. Therefore, the MSE lower bound for the position parameter 
$\boldsymbol{p}_{T}$ and synchronization parameter $\boldsymbol{\xi}_{b}$, based on the sensing receive signal 
$\tilde{\boldsymbol{y}}_{b}$, is expressed as {follows:}
\begin{small}
\begin{align}
&\mathrm{Cov}_{\tilde{\boldsymbol{y}}_{b}}((\hat{\boldsymbol{\zeta}}_{b}-\boldsymbol{\zeta}_{b})(\hat{\boldsymbol{\zeta}}_{b}-\boldsymbol{\zeta}_{b})^{T})\succeq\underbrace{\mathbb{E}\left\{\left(\frac{\partial\log p(\tilde{\boldsymbol{y}}_{b};\boldsymbol{\zeta}_{b})}{\partial \boldsymbol{\zeta}_{b}}\right)\right.}_{\text{hybrid Fisher information matrix}}\nonumber\\
&\underbrace{\left.\left(\frac{\partial\log p(\tilde{\boldsymbol{y}}_{b};\boldsymbol{\zeta}_{b})}{\partial \boldsymbol{\zeta}_{b}}\right)^{T}\right\}}_{\text{hybrid FIM}}.\label{pro16}
\end{align}
\end{small}%
The HFIM can be divided into the observed FIM (OFIM) and the prior FIM (PFIM), and the expression is given as {follows:}
\begin{small}
\begin{align}
&\mathbb{E}\left\{\left(\frac{\partial\log p(\tilde{\boldsymbol{y}}_{b};\boldsymbol{\zeta}_{b})}{\partial \boldsymbol{\zeta}_{b}}\right)\left(\frac{\partial\log p(\tilde{\boldsymbol{y}}_{b};\boldsymbol{\zeta}_{b})}{\partial \boldsymbol{\zeta}_{b}}\right)^{T}\right\}=\nonumber\\
&\underbrace{-\mathbb{E}_{\tilde{\boldsymbol{y}}_{b}|\boldsymbol{\zeta}_{b}}\left\{\left(\frac{\partial^{2}\log p(\tilde{\boldsymbol{y}}_{b}|\boldsymbol{\zeta}_{b})}{\partial \boldsymbol{\zeta}_{b}\partial \boldsymbol{\zeta}_{b}^{T}}\right)\right\}}_{\text{OFIM}~\boldsymbol{\Xi}_{O}}\underbrace{-\mathbb{E}_{\boldsymbol{\zeta}_{b}}\left\{\left(\frac{\partial^{2}\log p(\Delta\boldsymbol{\xi}_{b})}{\partial \boldsymbol{\zeta}_{b}\partial \boldsymbol{\zeta}_{b}^{T}}\right)\right\}}_{\text{PFIM}~\boldsymbol{\Xi}_{P}},\label{pro17}
\end{align}
\end{small}%
in which $p(\tilde{\boldsymbol{y}}_{b};\boldsymbol{\zeta}_{b})$ and $p(\Delta\boldsymbol{\xi}_{b})$ are expressed as
\begin{small}
\begin{align}
p(\tilde{\boldsymbol{y}}_{b};\boldsymbol{\zeta}_{b})=\left(1/\left(2\pi \sigma_{b}^{2L_{r}\bar{S}}\right)^{L_{r}/2}\right)e^{-\frac
{1}{2\sigma_{b}^{2}}(\tilde{\boldsymbol{y}}_{b}-\tilde{\boldsymbol{x}}_{b})^{T}(\tilde{\boldsymbol{y}}_{b}-\tilde{\boldsymbol{x}}_{b})}\label{pro18}
\end{align}
\end{small}%
and
\begin{small}
\begin{align}
p(\Delta\boldsymbol{\xi}_{b})=\left(1/(2\pi\sigma_{\xi}^{2M})^{M/2}\right)e^{-\sigma_{\xi}^{2}/2\Delta\boldsymbol{\xi}_{b}^{T}\Delta\boldsymbol{\xi}_{b}}.\label{pro19}
\end{align}
\end{small}%
Then, the HCRLB matrix corresponding to the sensing receiver at the 
$b$-th BS can be expressed as
\begin{small}
\begin{align}
\boldsymbol{{\mathrm{HCRLB}}}_{b}(\boldsymbol{\zeta}_{b})=[\mathrm{HFIM}_{b}(\boldsymbol{\zeta}_{b})]^{-1}.\label{pro20}
\end{align}
\end{small}%
%分析得知，OFIM 和 PFIM 分别表示观测数据和先验知识对估计误差界限的贡献。因此，与 CRLB 相比，HCRBB 加入了额外的先验 Fisher 信息，可以更准确地确定估计误差的下限，从而提供更精确的性能评估。为了简化用于估计目标位置 $\boldsymbol{p}$ 的 HCRLB 的推导，我们将用于估计
%$\boldsymbol{q}_{n}$ 的混合 FIM 分为多个子块
Based on the analysis, the OFIM and the PFIM represent the contributions of the observed data and prior knowledge to the estimation error bounds, respectively. Therefore, compared to the CRLB, the HCRLB incorporates additional prior Fisher information, which enables a more accurate determination of the lower bound of the estimation error, thereby providing a more precise performance evaluation. To simplify the derivation of the HCRLB for estimating the target position $\boldsymbol{\xi}_{b}$, we divide the hybrid FIM used for estimating 
$\boldsymbol{\zeta}_{b}$ into multiple sub-blocks
\begin{small}
\begin{align}
\mathrm{HFIM}_{b}(\boldsymbol{\zeta}_{b})=\left[\begin{matrix}
\frac{\partial^{2}\log p(\tilde{\boldsymbol{y}}_{b};\boldsymbol{\zeta}_{b})}{\partial\boldsymbol{p}\partial\boldsymbol{p}}&\frac{\partial^{2}\log p(\tilde{\boldsymbol{y}}_{b};\boldsymbol{\zeta}_{b})}{\partial\boldsymbol{p}\partial\Delta\boldsymbol{\xi}_{b}}\\
\left(\frac{\partial^{2}\log p(\tilde{\boldsymbol{y}}_{b};\boldsymbol{\zeta}_{b})}{\partial\boldsymbol{p}\partial\Delta\boldsymbol{\xi}_{b}}\right)^{T}&\frac{\partial^{2}\log p(\tilde{\boldsymbol{y}}_{b};\boldsymbol{\zeta}_{b})}{\partial\Delta\boldsymbol{\xi}_{b}\partial\Delta\boldsymbol{\xi}_{b}}
\end{matrix}\right],\label{pro21}
\end{align}    
\end{small}%
where the elements of $\frac{\partial^{2}\log p(\tilde{\boldsymbol{y}}_{b};\boldsymbol{\zeta}_{b})}{\partial\boldsymbol{p}\partial\boldsymbol{p}}$, $\frac{\partial^{2}\log p(\tilde{\boldsymbol{y}}_{b};\boldsymbol{\zeta}_{b})}{\partial\boldsymbol{p}\partial\Delta\boldsymbol{\xi}_{b}}$
, and $\frac{\partial^{2}\log p(\tilde{\boldsymbol{y}}_{b};\boldsymbol{\zeta}_{b})}{\partial\Delta\boldsymbol{\xi}_{b}\partial\Delta\boldsymbol{\xi}_{b}}$
are derived in the \textbf{Appendix~\ref{appB}}. It is worth noting that according to the derivations in the \textbf{Appendix~\ref{appB}}, each element of $\mathrm{HFIM}_{b}(\boldsymbol{\zeta}_{b})$ is a quadratic function of $\boldsymbol{w}=[\boldsymbol{w}_{1}^{H},\cdots,\boldsymbol{w}_{B}^{H}]^{H}$. Then, the HCRLB matrix of target position $\boldsymbol{p}$ corresponding to the $b$-th BS sensing receiver is calculated as
\begin{small}
\begin{align}
&\boldsymbol{{\mathrm{HCRLB}}}_{b}(\boldsymbol{p})=(\underbrace{\frac{\partial^{2}\log p(\tilde{\boldsymbol{y}}_{b};\boldsymbol{\zeta}_{b})}{\partial\boldsymbol{p}\partial\boldsymbol{p}}}_{\boldsymbol{\Xi}^{b}_{\boldsymbol{p}\boldsymbol{p}}}-\underbrace{\frac{\partial^{2}\log p(\tilde{\boldsymbol{y}}_{b};\boldsymbol{\zeta}_{b})}{\partial\boldsymbol{p}\partial\Delta\boldsymbol{\xi}_{b}}}_{\boldsymbol{\Xi}^{b}_{\boldsymbol{p}\Delta\boldsymbol{\xi}_{b}}}\nonumber\\
&(\underbrace{\frac{\partial^{2}\log p(\tilde{\boldsymbol{y}}_{b};\boldsymbol{\zeta}_{b})}{\partial\Delta\boldsymbol{\xi}_{b}\partial\Delta\boldsymbol{\xi}_{b}}}_{\boldsymbol{\Xi}^{b}_{\Delta\boldsymbol{\xi}_{b}\Delta\boldsymbol{\xi}_{b}}})^{-1}(\frac{\partial^{2}\log p(\tilde{\boldsymbol{y}}_{b};\boldsymbol{\zeta}_{b})}{\partial\boldsymbol{p}\partial\Delta\boldsymbol{\xi}_{b}})^{T})^{-1}.\label{pro22}
\end{align}    
\end{small}%
To {rigorously investigate an} impact of TS errors on target position estimation, we apply the matrix inversion \textbf{Theorem~2}\cite{b30} to expand equation (\ref{pro17}
) into equation (\ref{pro19}) at the top of the next page. It is well known that the CRLB is asymptotically tight under certain conditions, which is typically achieved through maximum likelihood estimation when the {signal-to-noise ratio (SNR)} and/or observation time are sufficiently large. However, the HCRLB is not always asymptotically tight, even though it may approach the optimal lower bound under certain conditions. Before using the HCRLB to assess estimation performance, its asymptotic tightness must be verified, ensuring the existence of an unbiased estimator whose MSE equals the HCRLB\cite{b8}. 
%To this end, we present the following lemma 
%$\mathbb{E}_{\boldsymbol{q}_{n}}\{\bar{\boldsymbol{F}}_{n}^{D}(\boldsymbol{q}_{n})\}=\mathbb{E}_{\Delta\boldsymbol{u}_{n}}\{\bar{\boldsymbol{F}}_{n}^{D}(\boldsymbol{q}_{n})\}$. Additionally, based on the derivations in (53)-(58) in the Appendix, it is observed that the elements in $\bar{\boldsymbol{F}}_{n}^{D}(\boldsymbol{q}_{n})$ are not functions of $\Delta\boldsymbol{u}_{n}$, thus yielding $\mathbb{E}_{\Delta\boldsymbol{u}_{n}}\{\bar{\boldsymbol{F}}_{n}^{D}(\boldsymbol{q}_{n})\}=\bar{\boldsymbol{F}}_{n}^{D}(\boldsymbol{q}_{n})$. 
Consequently, according to \cite{b32}, the HCRLB is given in (\ref{pro24}) at the top of the next page and the asymptotically tight {is proved} in \cite{b30}.
\begin{figure*}
\begin{small}
\begin{align}
\boldsymbol{{\mathrm{HCRLB}}}_{b}(\boldsymbol{p})=\underbrace{(\boldsymbol{\Xi}_{\boldsymbol{p}\boldsymbol{p}}^{b})^{-1}}_{\text{CRLB with target position}}+\underbrace{(\boldsymbol{\Xi}_{\boldsymbol{p}\boldsymbol{p}}^{b})^{-1}\boldsymbol{\Xi}_{\boldsymbol{p}\Delta\boldsymbol{\xi}_{b}}^{b}(\boldsymbol{\Xi}_{\Delta\boldsymbol{\xi}_{b}\Delta\boldsymbol{u}_{b}}^{b}-(\boldsymbol{\Xi}_{\boldsymbol{p}\Delta\boldsymbol{\xi}_{b}}^{b})^{T}(\boldsymbol{\Xi}_{\boldsymbol{p}\boldsymbol{p}}^{b})^{-1}\boldsymbol{F}_{\boldsymbol{p}\Delta\boldsymbol{\xi}_{b}}^{b})^{-1}(\boldsymbol{\Xi}_{\boldsymbol{p}\Delta\boldsymbol{u}_{b}}^{b})^{T}(\boldsymbol{\Xi}_{\boldsymbol{p}\boldsymbol{p}}^{})^{-1}}_{\text{Extra component due to TS errors}}.\label{pro24}  
\end{align}    
\end{small}
\hrulefill
\end{figure*}
%Since the asymptotic tightness of the H-CRLB is proven, it can be adopted to assess the parameter estimation error performance. To facilitate comparison, based on (19), the CRLB matrix of $\boldsymbol{p}$ at BS sensing receiver $n$ without TS errors is given by
%\begin{align}
%\mathrm{CRLB}_{n}(\boldsymbol{p})=(\boldsymbol{F}_{\boldsymbol{p}\boldsymbol{p}}^{n})^{-1}
%\end{align}
 %From (19) and (20), it is observed that the existence of TS errors results in generating additional increments in the H-CRLB of target position p.This is because TS errors can cause ambiguity in the measurement of signal propagation
% delay, resulting in the deviation in target position estimation.
 %Due to the significant impact of TS errors on position estimation, the beamforming of BS ISAC transmitters is designed based on the CRLB constraints with TS errors in Section IV. Furthermore,we design the coordinated beamforming based on the CRLB constraints under perfect TS in Section V for comparison.

%In this section, the CRLB-constrained sum-rate maximization problem is formulated for the C-ISAC system in the presence of CSI and TS errors. Then, we focus on tackling the resulting problem by invoking the penalty-based framework.

The HCRLB matrix represents the minimum variance of an unbiased estimator for a parameter, with its trace providing a lower bound on overall estimation performance. In a communication and sensing integration system, the goal is to minimize power through coordinated beamforming at the BS while accounting for TS errors. The HCRLB matrix is used to evaluate target estimation performance and as a constraint to balance sensing and communication within the power budget. The optimization problem is framed as a robust power minimization problem, where the HCRLB constraint limits sensing error and the power allocation meets the worst-case communication rate constraint. Thus, the problem is given by
\begin{subequations}
\begin{small}
\begin{align}
\min_{\boldsymbol{W}_{b},\tilde{\boldsymbol{t}}_{b}, \tilde{\boldsymbol{r}}_{b}, c_{b},c_{b,u}}&~\sum\nolimits_{b=1}^{B}\|\boldsymbol{W}_{b}\|^{2},\label{pro25a}\\
\mbox{s.t.}~
&\mathrm{tr}\{\boldsymbol{{\mathrm{HCRLB}}}_{b}(\boldsymbol{p})\}\leq\gamma_{b},~\forall~b\in\mathcal{B}&\label{pro25b}\\
&C(\boldsymbol{w}_{b,u},\boldsymbol{w}_{b,u^{\prime}})\geq(2^{\gamma_{u}}-1),&\label{pro25c}\\
&c_{b,u}=c_{b}\in\{0,1\}, \forall~b\in\mathcal{B},u\in\mathcal{U},&\label{pro25d}\\
&\tilde{\boldsymbol{t}}_{b}\in\mathcal{C}_{t,b},\tilde{\boldsymbol{r}}_{b}\in\mathcal{C}_{r,b}&\label{pro25e}\\
&\|\boldsymbol{t}_{b}^{n_{1}}-\boldsymbol{t}_{b}^{n_{2}}\|_{2}\geq D,&\label{pro25f}\\
&\|\boldsymbol{r}_{b}^{m_{1}}-\boldsymbol{r}_{b}^{m_{2}}\|_{2}\geq D,&\label{pro25g}
\end{align}\label{pro25}%
\end{small}%
\end{subequations}%
where $\gamma_{b}$ is the sensing accuracy constraint of BS sensing receiver $b$. It
is observed that problem (\ref{pro25b}) is non-convex, which is difficult
to tackle due to the {transmit power minimization problem}. {Moreover}, communication constraint 
(\ref{pro25c}) is non-convex. (\ref{pro25d}) is the BS selection constraint. (\ref{pro25e}) is the movable region constraint of the MAs. (\ref{pro25f}) and (\ref{pro25g}) are the minimum spacing constraints for MAs.

\section{Proposed CDRL Algorithm}\label{IV}
This paper models the optimization problem of the C-ISAC system as a constrained Markov decision process (CMDP), considering both communication and sensing channels. In {a} time slot $n$, each BS selects an action based on the observed state. After the action, the BS receives feedback in the form of a reward and a cost, and the environment {transits} to the next state with a certain probability. This process repeats as the BS observes a new state and makes decisions based on it. The following sections will describe the key components of the proposed CMDP model.
%本文通过将 C-ISAC 系统的优化问题建模为约束马尔可夫决策过程 (CMDP)，解决了通信和传感信道的特征。在每个时隙 $n$ 中，每个 BS 根据观察到的环境状态选择一个动作。选择后，BS 会以奖励和与所选动作相对应的成本的形式收到反馈。然后，环境以一定的概率转换到下一个状态。当 BS 在每个时隙中观察到新状态并根据当前状态做出决定时，此过程将继续。以下各节将详细描述所提出的 CMDP 模型的基本组成部分。

\subsection{Proposed CMDP Model}

\textbf{State Space:} The state observed by the BS can be represented as 
$\boldsymbol{s}^{(n)}=\{\hat{\boldsymbol{h}}_{b,u},\boldsymbol{H}_{b^{\prime},b}|b\in\mathcal{B},u\in\mathcal{U}\}$, which includes the acquired communication and sensing channels.

\textbf{Action Space:} Since the environment typically does not undergo drastic changes between consecutive time slots, we set action space as 
$\boldsymbol{a}^{(n)}=\{\mathrm{Re}(\boldsymbol{W}_{b}),\mathrm{Im}(\boldsymbol{W}_{b}),\tilde{\boldsymbol{t}}_{b}, \tilde{\boldsymbol{r}}_{b}, c_{b},c_{b,u}|b\in\mathcal{B},u\in\mathcal{U}\}$.

\textbf{Reward and cost functions:} In order to enable all BSs to minimize {transmit power} while satisfying the sensing and communication rate constraints, we {design} the instantaneous reward and cost functions as {follows:}
\begin{small}
\begin{align}
&r^{(n)}=-\sum\nolimits_{b=1}^{B}\|\boldsymbol{W}_{b}^{(n)}\|^{2}-\delta_{1}\sum_{b=1}^{B}\sum_{u=1}^{U}\prod_{b,u}(\text{if}~c_{b,u}=c_{b}, \prod_{b,u}=1~\nonumber\\
&\text{else}\prod_{b,u}=0)-\delta_{2}\sum_{b=1}^{B}\prod_{t,b}(\text{if}~\tilde{\boldsymbol{t}}_{b}\in\mathcal{C}_{t,b}, \prod_{b}=1~\text{else}~\prod_{b}=0)-\nonumber\\
&\delta_{3}\sum_{b=1}^{B}\prod_{r,b}(\text{if}~\tilde{\boldsymbol{r}}_{b}\in\mathcal{C}_{r,b}~\text{else}~\prod_{b}=0), \nonumber\\
&c^{(n)}=\mathrm{tr}\{\boldsymbol{{\mathrm{HCRLB}}}_{b}(\boldsymbol{p})\}^{(n)}-C(\boldsymbol{w}_{b,u},\boldsymbol{w}_{b,u^{\prime}})^{(n)}-\|\boldsymbol{t}_{b}^{n_{1}}-\boldsymbol{t}_{b}^{n_{2}}\|_{2}^{(n)}\nonumber\\
&-\|\boldsymbol{r}_{b}^{m_{1}}-\boldsymbol{r}_{b}^{m_{2}}\|_{2}^{(n)},\label{pro27}
\end{align}    
\end{small}%
where $\delta_{1}$, $\delta_{2}$, and $\delta_{3}$ are weighting factors that control the
influence of BS selection and movable region of MAs in the
overall reward. Following the strategy parameterized by $\mu$, the long-term
cumulative reward and cost starting from $n_{0}$ are
\begin{small}
\begin{align}
R(\mu)=\mathbb{E}_{\mu}\left[\sum\nolimits_{n=n_{0}}^{S}\gamma^{n-n_{0}}r^{(n)}\right], C(\mu)=\mathbb{E}_{\mu}\left[\sum\nolimits_{n=n_{0}}^{S}\gamma^{n-n_{0}}c^{(n)}\right],\label{pro28}
\end{align}    
\end{small}%
where $\gamma$ denotes the discounting factor. The objective of {the BS} in {this CMDP is}
\begin{small}
\begin{align}
\mu^{*}=\arg\min_{\mu}R(\mu),C(\mu)\leq\Gamma_{c}=\sum\nolimits_{n=n_{0}}^{T}-\gamma^{n-n_{0}}\eta_{c},\label{pro29}
\end{align}    
\end{small}%
where $\eta_{c}=(\gamma_{b}+2^{\gamma_{u}}-1+2D)$ is the threshold for SE in each subframe, corresponding to the constraint in (9a).

4) \textbf{Reward and cost functions:} In the $n$-th time slot, the state $\boldsymbol{s}^{(n+1)}=\{\boldsymbol{S}_{H}^{(n+1)}\}$ is evolved as $\boldsymbol{S}_{H}^{(n+1)}=\boldsymbol{S}_{H}^{(n)}$.

\subsection{Proposed Algorithm}
As the number of users and BSs grows, the action space expands exponentially, making training harder\cite{b33}. Traditional algorithms like DQN struggle with high-dimensional spaces and cost constraints. To tackle this, we use an actor-critic framework with primal-dual updates and Wolpertinger-based action selection.

To address this constrained optimization problem, we design reward and cost evaluation networks parameterized by $\eta_{1}$ and $\eta_{2}$, which map from the joint state and action spaces to the corresponding rewards and costs, thus serving the purpose of action evaluation: 
$Q_{1}(\boldsymbol{s}
^{(n)}, \boldsymbol{a}^{(n)}|\eta_{1})$ and 
$Q_{2}(\boldsymbol{s}
^{(t)}, \boldsymbol{a}^{(n)}|\eta_{2})$. The actor network $\kappa$ is parameterized by 
$\eta_{3}$, which maps the state space to actions 
$\kappa(\boldsymbol{s}^{(n)}|\eta_{3})=\boldsymbol{a}^{(n)}$. To reduce fluctuations in the target values during training and accelerate convergence, we employ target networks $Q_{1}^{\prime}$, $Q_{2}^{\prime}$, and $\kappa^{\prime}$, parameterized by 
$\eta_{1}^{\prime}$, 
$\eta_{2}^{\prime}$, and 
$\eta_{3}^{\prime}$, respectively. To solve the CMDP presented in (\ref{pro30}), we first apply the Lagrangian relaxation method to transform the problem in (\ref{pro29}) into an unconstrained {problem as follows:}
\begin{small}
\begin{align}
\min_{\zeta\geq 0}\max_{\eta_{3}}[R(\eta_{3})-\zeta(C(\eta_{3})-\Gamma_{c})],\label{pro30}
\end{align}
\end{small}%
{where} the dual variable $\zeta$ acts as an additional parameter that is updated during the optimization process. 

\textbf{Action Selection}: With $\kappa$ held constant, the BS {aims} to maximize the unconstrained objective in {Eq.(\ref{pro30})}. Consequently, one action $\boldsymbol{a}^{*}$
is chosen from the set of $K$ possible actions and executed by the BS at time slot $n$, based on the current feedback from the critic networks.
\begin{small}
\begin{align}
\boldsymbol{a}^{*(t)}=\arg\max_{\boldsymbol{a}\in\hat{\mathcal{A}}}[Q_{1}(\boldsymbol{s}^{(n)},\boldsymbol{a}|\eta_{1})-\lambda(Q_{2}(\boldsymbol{s}^{(n)},\boldsymbol{a}|\eta_{2})-\Gamma_{c})].\label{pro31}
\end{align}    
\end{small}%
The transition tuple $(\boldsymbol{s}^{(n)},\boldsymbol{a}^{(n)},r^{(n)},c^{(n)},\boldsymbol{s}^{(n+1)})$ is stored in a
memory replay buffer $\mathcal{M}$ for network updating.

\begin{figure}[htbp]
  \centering
  \includegraphics[scale=0.30]{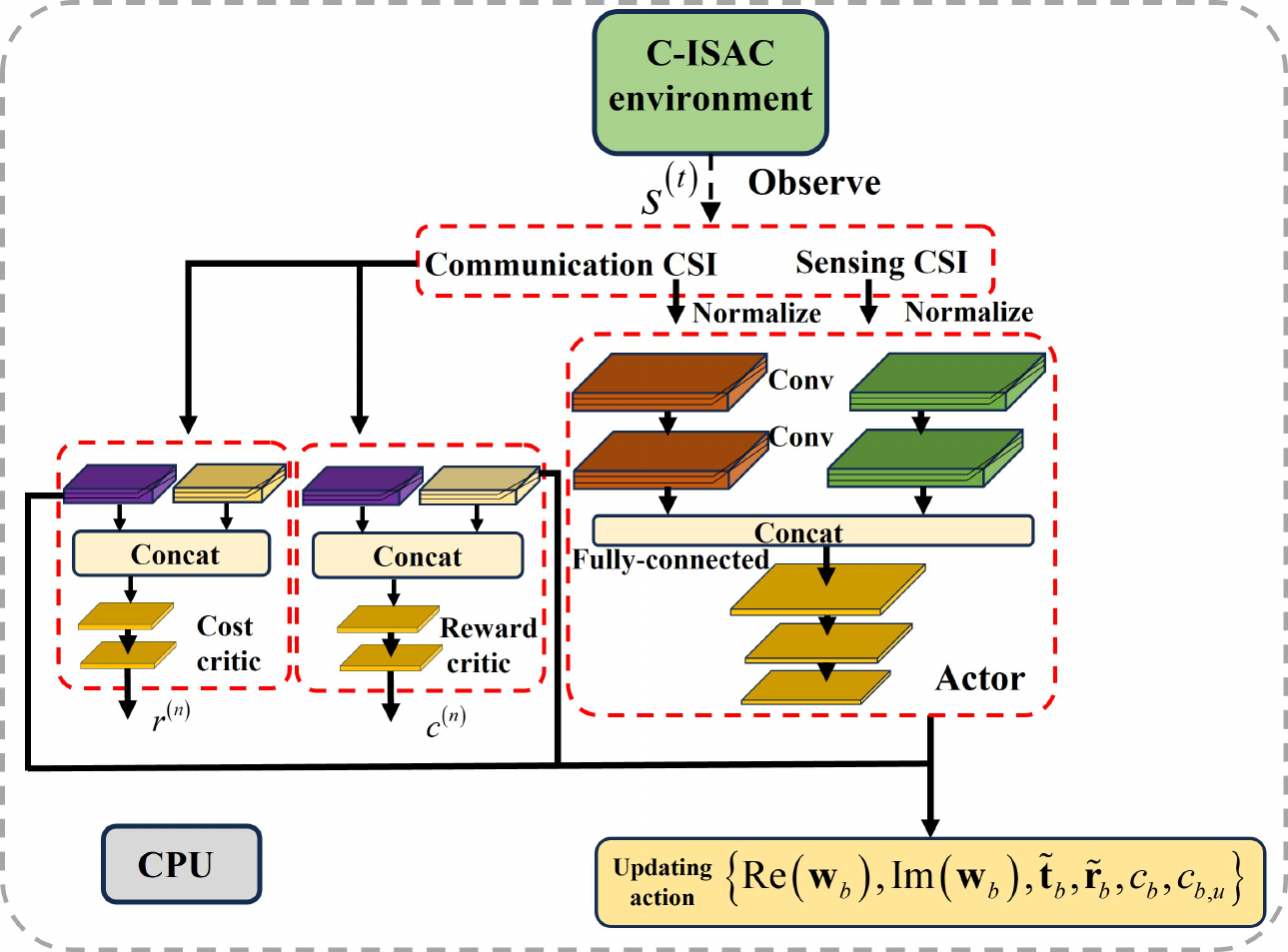}
  \captionsetup{justification=centering}
  \caption{The proposed robust algorithm based on CDRL.}\vspace{-10pt}
\label{FIGURE2}
\end{figure}

\textbf{Primal-Dual Network Update}: To facilitate the model
training, we adopt a primal dual-deep deterministic policy gradient (PD-DDPG) {method}, where the policy parameterized
by $\eta_{3}$ and the dual variable $\lambda$ are updated alternately for
objective maximization while the critics $\eta_{1}$ and $\eta_{2}$ are updated
for more precise policy evaluation. To be specific, a batch of
Nb transition tuples are randomly sampled from $\mathcal{M}$, and the
target {reward $y_i$} and target {cost $z_i$ are} calculated as
\begin{small}
\begin{align}
&y_{i}=r_{i}+\gamma Q_{1}^{\prime}(\boldsymbol{s}_{i+1},\kappa^{\prime}(\boldsymbol{s}_{i+1}|\eta_{3}^{\prime})|\eta_{1}^{\prime}),\nonumber\\
&z_{i}=c_{i}+\gamma Q_{2}^{\prime}(\boldsymbol{s}_{i+1},\kappa^{\prime}(\boldsymbol{s}_{i+1}|\eta_{3}^{\prime})|\eta_{2}^{\prime}),\label{pro32}
\end{align}    
\end{small}%
where $i=1,\cdots, N_{b}$. Then the reward and cost critic networks are updated by minimizing the mean square error
\begin{small}
\begin{align}
&f_{1}=\sum\nolimits_{i=1}^{N_{b}}(y_{i}-Q_{1}(\boldsymbol{s}_{i},\boldsymbol{a}_{i}|\eta_{1}))^{2}/N_{b},\nonumber\\
&f_{2}=\sum\nolimits_{i=1}^{N_{b}}(z_{i}-Q_{2}(\boldsymbol{s}_{i},\boldsymbol{a}_{i}|\eta_{2}))^{2}/N_{b},\label{pro33}
\end{align}    
\end{small}%
with the learning {rates} $\kappa_{1}$ and $\kappa_{2}$ respectively, followed by the actor updating through the sampled policy gradient ascend as
\begin{small}
\begin{align}
\eta_{3}^{(n+1)}=\eta_{3}^{(n)}+\kappa_{3}\Delta_{\eta_{3}}[R(\eta_{3})-\kappa(C(\eta_{3})-\Gamma_{c})],\label{pro34}
\end{align}
\end{small}%
with the gradient being $\sum_{i}\Delta_{\eta_{1}}[Q_{1}(\boldsymbol{s}_{i},\boldsymbol{a}_{i}|\eta_{1})-\kappa Q_{2}(\boldsymbol{s}_{i},\boldsymbol{a}_{i}|\eta_{2})]|_{\boldsymbol{s}=\boldsymbol{s}_{i}}/N_{b}$ and the learning rate as $\kappa_{1}$. Next, the dual variable is updated by gradient descent as
\begin{small}
\begin{align}
\kappa^{(n+1)}=\max(\kappa^{(n)}+\rho_{\kappa}\Delta_{\kappa},0),\label{pro35}
\end{align}    
\end{small}%
with {$\Delta_{\kappa}=\sum_{i=1}^{N_{b}}[Q_{2}(\boldsymbol{s}_{i},\kappa(\boldsymbol{s}_{i}|\eta_{1}))-\Gamma_{c}]/N_{b}$} and the step size
$\rho_{\kappa}$. Finally, we perform the soft updating of the target networks
as\cite{b34}. The pseudo-code of the training stage of the proposed
scheme is illustrated in Alg. 1. After completing the offline
training, the proposed scheme applies the trained actor to
perform online updates of $\{\mathrm{Re}(\boldsymbol{W}_{b}),\mathrm{Re}(\boldsymbol{W}_{b}),\tilde{\boldsymbol{t}}_{b},\tilde{\boldsymbol{r}}_{b},c_{b},c_{b,u}\}$. The framework of the CDRL is demonstrated in \textbf{Fig}.\ref{FIGURE2}.
Finally, the proposed algorithm is summarized in \textbf{Algorithm~\ref{algo1}}.
\begin{algorithm}%
\caption{Proposed CDRL algorithm for (\ref{pro25})} \label{algo1}%算法的名字
\hspace*{0.02in}{\bf Initialize:}
$\kappa(\boldsymbol{s}|\eta_{1})$, $Q_{1}(\boldsymbol{s},\boldsymbol{a}|\eta_{2})$, $Q_{2}(\boldsymbol{s},\boldsymbol{a}|\eta_{2})$, $\eta_{1}^{\prime}=\eta_{1}$,$\eta_{2}^{\prime}=\eta_{2}$,$\eta_{2}^{\prime}=\eta_{2}$, $\kappa=0$, $\mathcal{M}=\emptyset$, $N_{b}=64$, $\gamma=0.5$, random Ornstein-Uhlenbeck process $\mathcal{N}$ and initial state $\boldsymbol{s}^{(0)}$.\\
\hspace*{0.02in}{\bf Repeat:}~$n=n+1$.\\
$\boldsymbol{a}^{(n)}=\kappa(\boldsymbol{s}^{(n)}|\eta_{1})$.\\
$\boldsymbol{a}^{(n)}$ is selected based on (\ref{pro32}).\\
Computing $r^{(n)}$, $c_{1}^{(n)}$ and $\boldsymbol{s}^{(n+1)}$\\
Store the transition $(\boldsymbol{s}^{(n)}, \boldsymbol{a}^{(n)}, r^{(n)}, c^{(n)}, \boldsymbol{s}^{(n+1)})$ in $\mathcal{M}$\\
Sample a random batch of $N_{b}$ transitions from$\mathcal{M}$\\
Computing (\ref{pro32}), (\ref{pro33}), (\ref{pro34}) and (\ref{pro35}).\\
Updating the network parameters.
\end{algorithm}

\subsection{Computation Complexity Analysis}
Denote $N_{1}$ as the maximal number of convolutional layers, $N_{2}$ as the maximal
number of input and output feature maps, and $N_{3}$ as the
maximal side length of the filters in convolutional neural networks (CNN). $N_{4}$ and $N_{5}$ as
the maximal number of neurons in the hidden layers and the
number of hidden layers in fully-connected networks(FCN). Then the complexity in the
forward pass of the actor-network in online deployment is
$\mathcal{O}(N_{1}(NBU+NMB)N_{3}^{2}N_{2}^{2}+N_{5}^{2}N_{4})$\cite{b35,b36}.

\section{Numerical Results}\label{V}
\begin{figure}[htbp]
  \centering
  \includegraphics[scale=0.25]{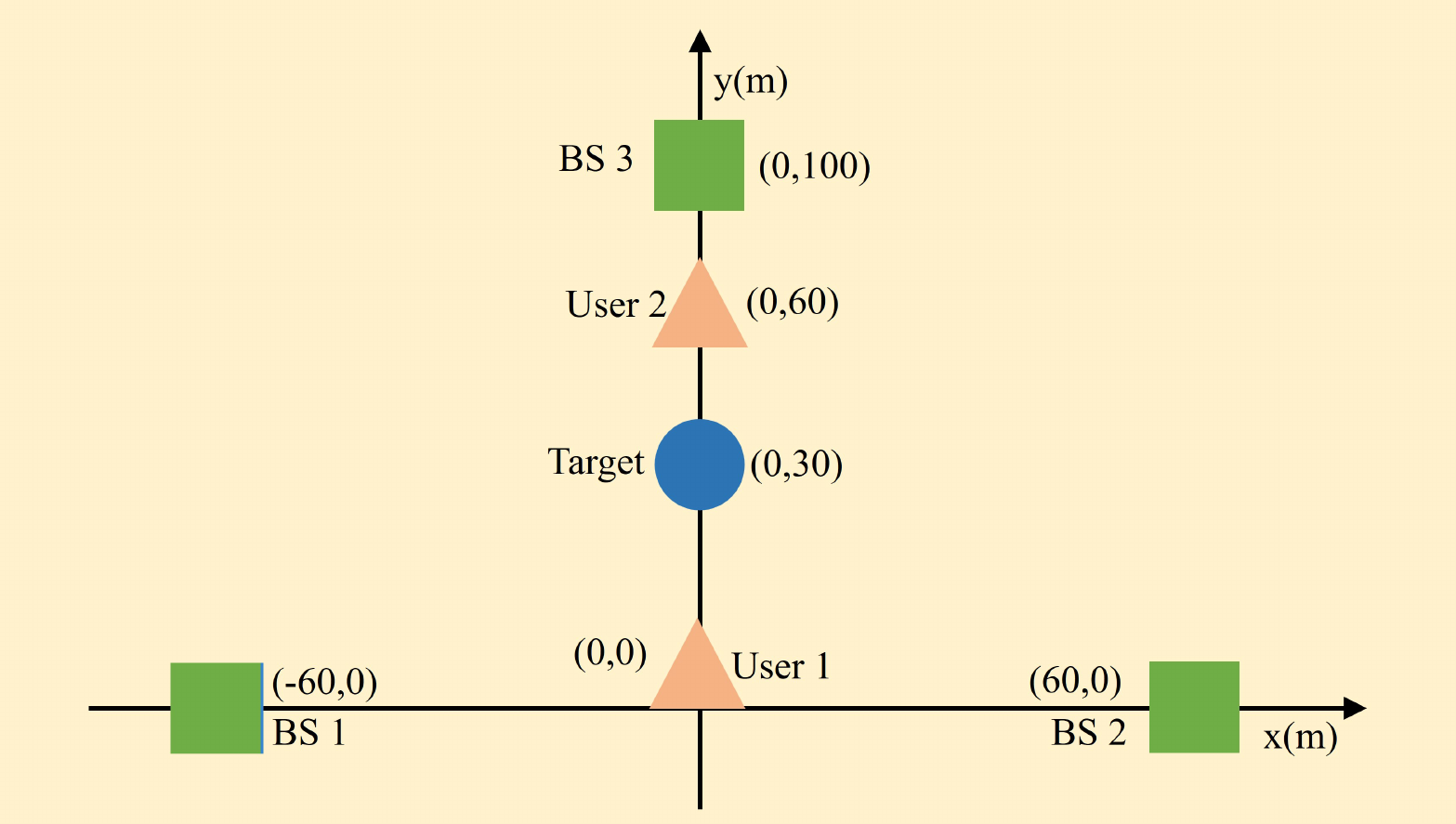}
  \captionsetup{justification=centering}
  \caption{Simulation setup of the MA-enabled C-ISAC system.}\vspace{-10pt}
\label{FIGURE3}
\end{figure}
In the simulation, the positions of the BSs, users, and targets are shown in \textbf{Fig}.\ref{FIGURE3}, where $B=3$, $U=2$, and $T=1$. Each BS is equipped with $N=8$ MAs and $M = 4$ receive MAs, where $N_{x} = 4$, $N_{y} = 2$, $M_{x} = 2$, and $M_{y} = 2$. The channel path loss is modeled as $PL(d)=PL_{0}(d/d_{0})^{-\Omega}$, where $PL_{0}$ denotes the path loss factor, $d_{0}$ represents the reference distance, $d$ indicates the link distance, and $\Omega$ represents the path loss, with $PL_{0}=-30$~dB, $d_{0}=1$~m. The number of channel paths is $L_{b,u} = 3$. The noise power level is $-120$~dBm. The minimum distance between MAs is set to $D= \lambda/2$. The transmission range of the {MAs} is $x_{t,b}^{min} = -2\lambda$, $x_{t,b}^{max} = 2\lambda$, and $y_{t,b}^{min} = -2\lambda$. The range for the receive {MAs} is $x_{r,b}^{min} = -2\lambda$, $x_{r,b}^{max} = 2\lambda$, and $y_{r,b}^{min} = -2\lambda$, $y_{r,b}^{max} = 2\lambda$. The path loss exponent for the BS-user link is set to $2.8$, while the path loss exponent for the BS-target link is set to $2.2$. The target's radar cross section (RCS) is given as $\alpha = 10$. Sensing accuracy $\gamma_{b}=0.05$, and communication rate $\gamma_{b}=2$ bit/s/Hz
%{}{In addition to the rate, the energy efficiency}, which is defined as the ratio of sum-rate to total consumption power, that is,
%\begin{eqnarray}
%EE=\frac{R}{P+N_{RF}P_{RF}+N_{PS}P_{PS}+N_{r}P_{RIS}},\label{3-68}
%\end{eqnarray}
%where $P_{RF}$ is the power consumption of each RF chian, and $N_{RF}=N{t}$ is an all-digital structure, and when $N_{RF}=N$ is a mixed structure. 
%$P_{PS}$ is the power consumed by each phase shift, and the number of phase shifters is also $N_{PS}$, where it is an all-digital structure when $N_{PS}=0$ and a mixed structure when $N_{PS}=NN_{T}$. $P_{RIS}$ is the power consumed by each reflecting elements. 
%In the simulation, $P_{RF}=300$~mW and $P_{PS}=40$~mW\cite{dai2018hybrid}, $P_{RIS}=10$~mW\cite{huang2019reconfigurable}.

%\begin{figure}[!h]
%\centering
%\includegraphics[scale=0.45]{FIG3_2E.pdf}
%  \vspace{2.0cm}
%  \medskip
  %\caption{Simulated RIS-aided mmWave-NOMA communication scenario in (a). Simulated mmWave-NOMA communication scenario without RIS in (b)\label{fig:3_2}.}
%\end{figure}
\begin{figure}[htbp]
\centering
\begin{minipage}[t]{0.48\textwidth}
\centering
\includegraphics[width=0.82\textwidth, height=0.48\textwidth]{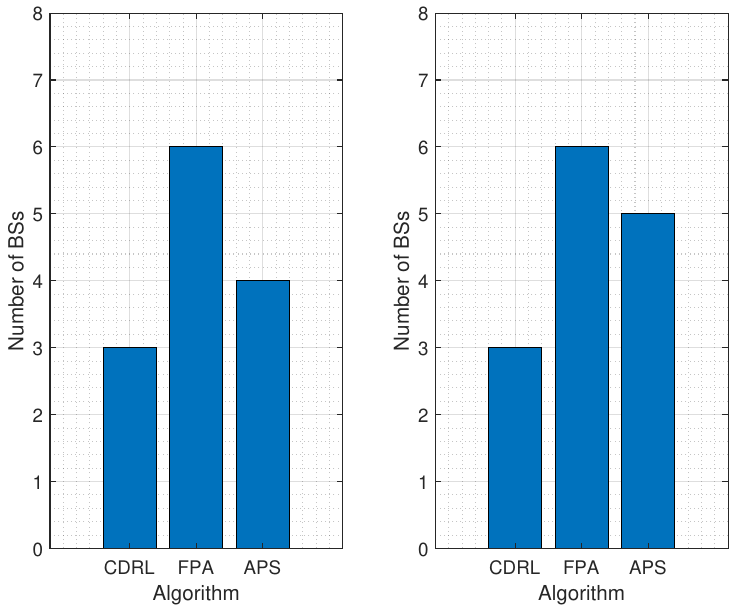}
\put(-150,-5){\small\textbf{(a)}}
\put(-60,-5){\small\textbf{(b)}}
\caption{(a) Cumulative reward of the CDRL versus epochs. (b) Cumulative cost of the CDRL versus epochs.}\vspace{-10pt}
\label{FIGURE4}
\end{minipage}
\end{figure}

We validate the proposed scheme in \textbf{Fig}.\ref{FIGURE4} by examining the cumulative reward and cost achieved by the agent during training. In the experiment, we set the TS error to $\sigma_{\xi} = 100$ ns and $\bar{\epsilon}_{b,u} = 0.01$. As {the training} progresses, the cumulative reward increases, indicating that the agent learns to optimize its strategy to meet the system’s goals. Meanwhile, the BS adjusts its behavior to reduce the cumulative cost while satisfying both the communication rate and sensing accuracy constraints. As training continues, the agent refines its decision-making, reducing the cumulative cost below the threshold, demonstrating that the agent can find a near-optimal solution without exhaustive search. By adjusting its policy, the agent balances reward and cost, optimizing performance under the system’s constraints. {As shown in \textbf{Fig}.\ref{FIGURE4}(a), despite the presence of TS and CSI errors, the reward values of the CDRL algorithm decrease slightly but remain close to those of the three alternative schemes. This indicates that, even in the intermediate stages before the policy fully converges, the algorithm can maintain a high level of stability and robustness. The reason for this lies in the robust optimization design of the CDRL algorithm, which incorporates worst-case communication rate constraints. This design enables the algorithm to perform well in the presence of environmental errors and effectively mitigates the negative impact of errors. Therefore, even when errors are present and the policy has not fully converged, the CDRL algorithm still demonstrates high robustness and stability.} Additionally, \textbf{Fig}.\ref{FIGURE4}(b) shows that TS and CSI errors cause a significant increase in cumulative cost and a decrease in cumulative reward. These errors require more resources to meet the constraints, increasing the number of episodes needed for convergence.

\begin{figure}[htbp]
\centering
\begin{minipage}[t]{0.48\textwidth}
\centering
\includegraphics[width=0.82\textwidth, height=0.48\textwidth]{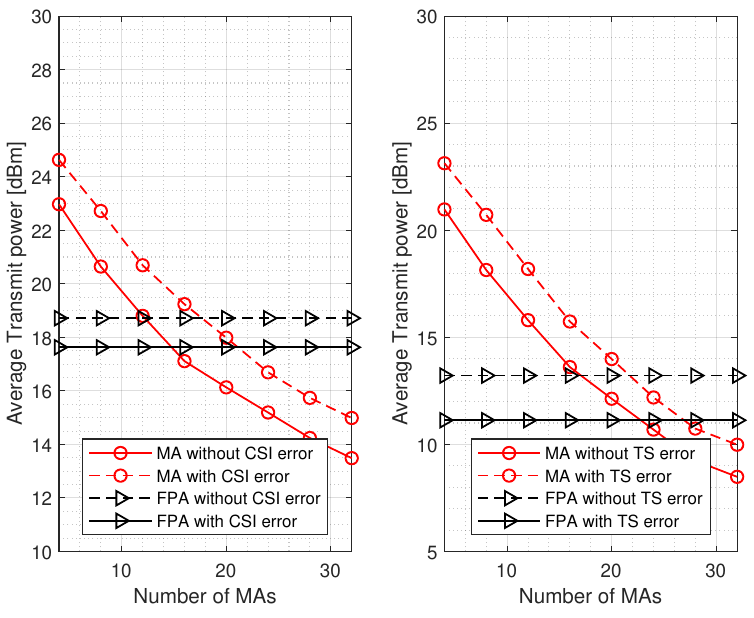}
\put(-150,-5){\small\textbf{(a)}}
\put(-60,-5){\small\textbf{(b)}}
\caption{(a) Average transmit power versus the number of MAs with CSI error. (b) Average transmit power versus the number of MAs with TS error.}\vspace{-10pt}
\label{FIGURE5}
\end{minipage}
\end{figure}

As shown in \textbf{Fig}.\ref{FIGURE5}, the simulation investigates the relationship between the number of MAs and the system's {transmit power}. The results indicate that as the number of MAs increases, the system's {transmit power} decreases. This is because additional {MAs} enhance the system's spatial diversity performance. By increasing the number of antennas, the simulation models the signal reception process under various channel environments and finds that multiple antennas at different locations receiving signals can effectively mitigate multipath effects. By adjusting the antenna positions and utilizing beamforming techniques, the system can enhance signal directionality and suppress interference, ensuring good communication quality and target tracking accuracy under lower {transmit power}. The simulation results show that under the same power consumption when there is a CSI error, only $16$ MAs are required to achieve an FPA of $32$. This indicates that the number of MAs can not only reduce power consumption but also lower the BS cost. When TS errors are present, only $20$ MAs are needed to achieve an FPA of $32$, demonstrating that a small number of MAs can effectively resist TS errors. The above simulation results suggest that as the number of MAs increases, the system can meet the communication robustness requirements with lower power consumption, achieving both energy savings and resource optimization.

\begin{figure}[htbp]
\centering
\begin{minipage}[t]{0.48\textwidth}
\centering
\includegraphics[width=0.82\textwidth, height=0.48\textwidth]{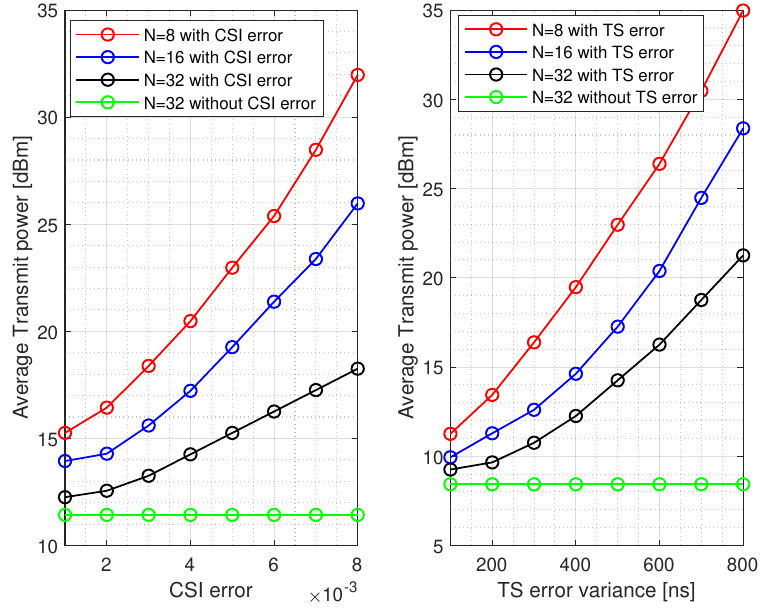}
\put(-150,-5){\small\textbf{(a)}}
\put(-60,-5){\small\textbf{(b)}}
\caption{(a) Average transmit power versus the CSI error. (b) Average transmit power versus the TS error.}\vspace{-10pt}
\label{FIGURE6}
\end{minipage}
\end{figure}
In \textbf{Fig}.\ref{FIGURE6}(a), we illustrate the relationship between the transmit power and the CSI error. It can be observed that as the CSI error increases, the transmit power of the system increases significantly. This is because CSI error reduces the accuracy of the channel state information, which in turn degrades the quality of the receive signal, leading to a decrease in SINR. To compensate for the reduction in SINR, the system must increase the transmit power to enhance the signal strength, thereby ensuring sufficient signal quality to overcome noise and interference, and maintaining the reliability of the communication link as well as the accuracy of target tracking. By increasing the transmit power, the system effectively improves SINR, reduces the estimation bias caused by CSI errors, and ensures the system can maintain high performance. In \textbf{Fig}.\ref{FIGURE6}(b), we demonstrate the relationship between the consumed power and the TS error under the same parameter settings as in \textbf{Fig}.\ref{FIGURE6}(a). Power consumption with a TS error is significantly higher than without, and the gap increases as the TS error grows. This is because TS errors force the system to consume more power to meet the HCRLB constraint, which minimizes estimation error for optimal performance. As TS error increases, uncertainty in the system’s estimation grows, requiring more power to maintain reliable and accurate tracking. The receive signal becomes more distorted, reducing the accuracy of target state estimation. To compensate for this loss in precision, the system invests additional power to boost signal strength, reducing estimation bias caused by TS errors and ensuring high performance under the HCRLB constraint. Therefore, as TS error increases, power consumption rises, reflecting the extra resources needed for high-precision tracking in the face of growing estimation errors.

\begin{figure}[htbp]
\centering
\begin{minipage}[t]{0.48\textwidth}
\centering
\includegraphics[width=0.82\textwidth, height=0.48\textwidth]{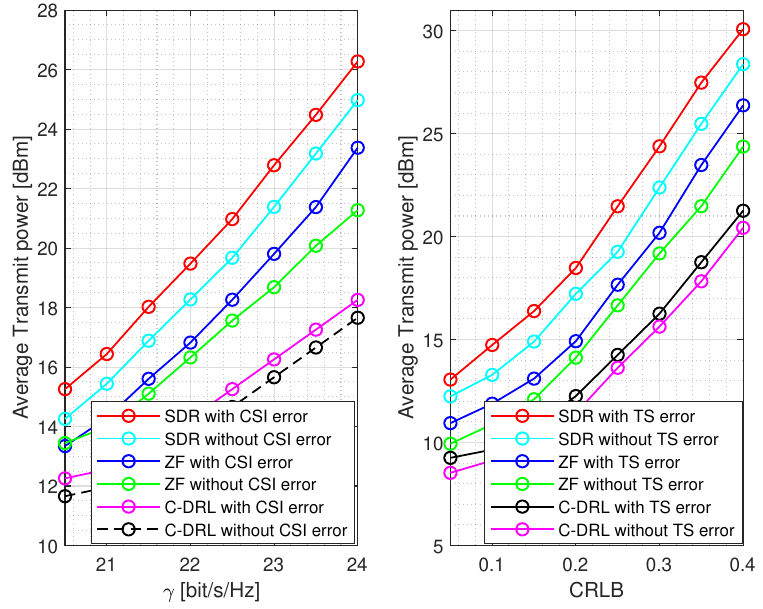}
\put(-150,-5){\small\textbf{(a)}}
\put(-60,-5){\small\textbf{(b)}}
\caption{(a) Average transmit power versus the communication rate constraint with proposed algorithm and nonrobust algorithms. (b) Average transmit power versus the HCRLB constraint with proposed algorithm and nonrobust algorithms.}\vspace{-10pt}
\label{FIGURE7}
\end{minipage}
\end{figure}
Finally, to further validate the performance of the proposed coordinated transmission beamforming algorithm under different communication rates, this section compares it with existing beamforming methods based on semidefinite relaxation (SDR) and zero-forcing (ZF). \textbf{Fig}.\ref{FIGURE7}(a) illustrates the relationship between {transmit power}, communication rate, and sensing accuracy under the conditions of $\sigma_{\xi} = 100$ ns and $\bar{\epsilon}_{b,u} = 0.01$. The simulation results show that as the communication rate increases, the system’s {transmit power} also increases. This is because higher communication rates require stronger signal strength to maintain a sufficient SINR that satisfies the communication demand. Additionally, the simulation results indicate that, under poor CSI conditions, the proposed robust design achieves better performance in increasing the communication rate while keeping power consumption lower compared to both the SDR and ZF-based methods.
To further investigate the relationship between {transmit power} and HCRLB constraints, \textbf{Fig}.\ref{FIGURE7}(b) shows the correlation between {transmit power} and HCRLB constraints. The results indicate that as the HCRLB constraint becomes more stringent, the system’s {transmit power} increases significantly. This is because a stricter HCRLB constraint requires more precise target state estimation, thus demanding higher power to meet the more stringent sensing accuracy requirements. Conversely, as the HCRLB constraint is relaxed, the system's {transmit power} decreases, while still being able to satisfy less strict estimation requirements with lower power. The simulation results further demonstrate that, in the presence of TS errors, the proposed robust design adapts better to different HCRLB constraints compared to the non-robust design, thereby reducing power consumption while maintaining other system performance. Furthermore, it outperforms both SDR- and ZF-based methods in terms of power efficiency and performance.

\begin{figure}[htbp]
\centering
\begin{minipage}[t]{0.48\textwidth}
\centering
\includegraphics[width=0.82\textwidth, height=0.48\textwidth]{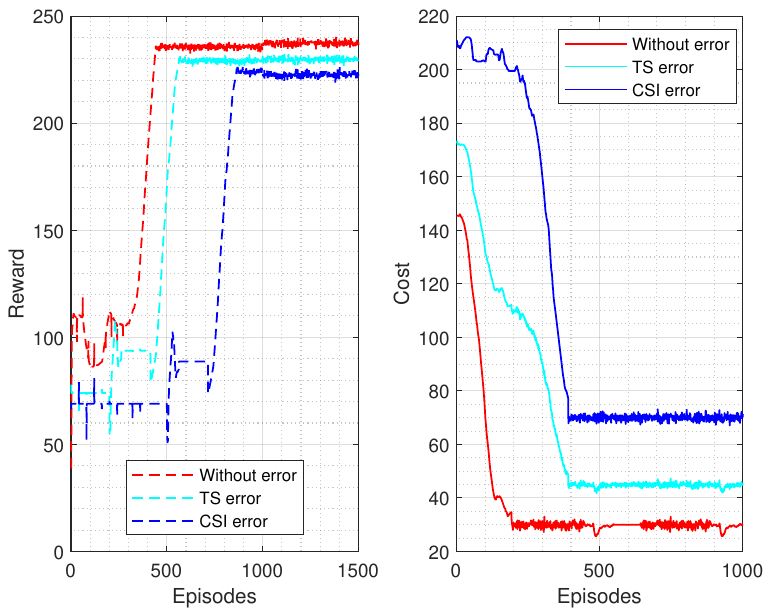}
\put(-150,-5){\small\textbf{(a)}}
\put(-60,-5){\small\textbf{(b)}}
\caption{(a) BS selection versus the proposed algorithm and existing algorithms with $\bar{\epsilon}_{b,u}=0.01$ and $\sigma_{\xi}=100$~ns. (b) BS selection versus the proposed algorithm and existing algorithms with $\bar{\epsilon}_{b,u}=0.01$ and $\sigma_{\xi}=200$~ns.}\vspace{-10pt}
\label{FIGURE8}
\end{minipage}
\end{figure}
Based on the results shown in simulation \textbf{Fig}.~\ref{FIGURE8}(a), this paper compares three schemes: the CDRL scheme, the Adaptive Portable Antenna (APS) scheme, and the FPA scheme, with a CSI error of $0.01$ in the simulation. The results show that, under a $100$~ns synchronization error, the CDRL scheme achieves the desired performance with only $3$ BSs, while the APS scheme needs $4$ BSs and the FPA scheme requires $6$ BSs. This demonstrates the CDRL scheme's strong adaptability and lower BS requirements, achieving the same performance with fewer resources. The CDRL scheme optimizes {MAs}, allowing flexible adjustment of antenna positions and beamforming, improving signal diversity and mitigating errors. In contrast, the APS scheme, though using {MAs}, is less flexible and requires more BSs to maintain accuracy. The FPA scheme, with fixed antennas, cannot adjust dynamically and needs more BSs to ensure performance under error conditions. In \textbf{Fig}.~\ref{FIGURE8}.(b), when the synchronization error increases to $200$~ns, the CDRL scheme still only requires $3$ BSs, while the APS scheme requires $5$ BSs, and the FPA scheme still requires $6$ BSs. This confirms the superiority of the CDRL scheme, especially when synchronization errors are large. The CDRL scheme effectively mitigates the impact of these errors, keeping BS demand low. This is due to its flexibility, which leverages {MAs} to maintain low BS requirements under varying error conditions. In contrast, while the APS scheme also uses {MAs}, its error tolerance is weaker, requiring more BSs to compensate for performance loss. The FPA scheme, with its fixed antenna configuration, cannot adjust dynamically and relies on adding more BSs to maintain performance, especially under large errors. In conclusion, the CDRL scheme outperforms others by reducing BSs, cutting power consumption, and ensuring high robustness and efficiency in error-prone environments.

\section{Conclusion}\label{VI}
This paper investigates the performance of a C-ISAC network, where distributed BSs coordinate beamforming for user communication while simultaneously performing static target sensing. We {have addressed} the challenges introduced by imperfect CSI and TS errors, which degrade system performance. To improve robustness, we {have proposed} the use of MAs, which can enhance the system's resilience to these errors.
Through the analysis of CSI and TS errors, we {have derived} models for the worst-case achievable rate and sensing precision. To reduce power consumption, we {have optimized} beamforming, MA position, and BS selection, ensuring that the system meets the required performance under these errors. The optimization problem, being non-convex, {has been} tackled using a constrained C-DRL approach, incorporating a tailored DDPG algorithm with Wolpertinger architecture. Simulation results {have shown} that our proposed method significantly improves system robustness and outperforms traditional fixed-antenna systems. Future work will explore the integration of dynamic environmental factors, such as user mobility and varying interference conditions, into the optimization process. Additionally, the use of advanced machine learning algorithms to further optimize MA-aided C-ISAC systems will be investigated, aiming for even greater performance improvements in real-world scenarios.
%本文研究了 C-ISAC 网络的性能，其中分布式 BS 协调波束成形以进行用户通信，同时执行静态目标感知。我们解决了不完美的 CSI 和 TS 错误带来的挑战，这些错误会降低系统性能。为了提高稳健性，我们建议使用 MA，这可以增强系统对这些错误的恢复能力。通过对 CSI 和 TS 错误的分析，我们推导出最坏情况下可实现的速率和感知精度的模型。为了降低功耗，我们优化了波束成形、MA 位置和 BS 选择，确保系统在这些错误下满足所需的性能。优化问题是非凸的，使用约束 C-DRL 方法解决，结合了具有 Wolpertinger 架构的定制 DDPG 算法。仿真结果表明，我们提出的方法显著提高了系统的稳健性，并且优于传统的固定天线系统。未来的工作将探索将动态环境因素（例如用户移动性和不断变化的干扰条件）集成到优化过程中。此外，我们将研究使用先进的机器学习算法进一步优化 MA 辅助 C-ISAC 系统，旨在在现实场景中实现更大的性能提升。
%\footnote{We carefully studied some references, for example, ``Robust Beamforming Design for Intelligent Reflecting Surface Aided MISO Communication Systems" in IEEE Wireless Communications Letters and ``A Framework of Robust Transmission Design for IRS-aided MISO Communications with Imperfect Cascaded Channels" in IEEE Transactions Signal Processing. They are useful to improve our future work, and we will combine the robust design proposed in the two papers to solve the RIS-aided mmWave NOMA system with imperfect CSI. 
%}.
\begin{appendices}
\section{The proof of \textbf{Theorem}~\ref{th1}}\label{APP1}
The communication rate constraint in (\ref{pro12}) can be rewritten as
\begin{small}
\begin{align}
&\sum\nolimits_{b=1}^{B}|c_{b,u}(\hat{\boldsymbol{h}}_{b,u}(\tilde{\boldsymbol{t}}_{b})+\Delta\boldsymbol{h}_{b,u})^{H}\boldsymbol{w}_{b,u}|^{2}/\left(\sum\nolimits_{b=1}^{B}\sum\nolimits_{u^{\prime}=1,u^{\prime}\neq u}^{U}\right.\nonumber\\
&\left.\left.|c_{b,u^{\prime}}(\hat{\boldsymbol{h}}_{b,u^{\prime}}(\tilde{\boldsymbol{t}}_{b})+\Delta\boldsymbol{h}_{b,u^{\prime}})^{H}\boldsymbol{w}_{b,u^{\prime}}|^{2}\right)+\sigma_{b}^{2}\right)\geq 2^{\gamma_{u}}-1.\label{proA1}     
\end{align}
\end{small}
The constraint in (\ref{proA1}) is further expressed as
\begin{small}
\begin{align}
&\sum\nolimits_{b=1}^{B}|c_{b,u}(\hat{\boldsymbol{h}}_{b,u}(\tilde{\boldsymbol{t}}_{b})+\Delta\boldsymbol{h}_{b,u})^{H}\boldsymbol{w}_{b,u}|^{2}\geq (2^{\gamma_{u}}-1)\left(\sum\nolimits_{b=1}^{B}\right.\nonumber\\
&\left.\left.\sum\nolimits_{u^{\prime}=1,u^{\prime}\neq u}^{U}|c_{b,u^{\prime}}(\hat{\boldsymbol{h}}_{b,u^{\prime}}(\tilde{\boldsymbol{t}}_{b})+\Delta\boldsymbol{h}_{b,u^{\prime}})^{H}\boldsymbol{w}_{b,u^{\prime}}|^{2}\right)+\sigma_{b}^{2}\right).\label{proA2}  
\end{align}
\end{small}
According to the triangle inequality, we have 
\begin{small}
\begin{align}
&|c_{b,u}(\hat{\boldsymbol{h}}_{b,u}(\tilde{\boldsymbol{t}}_{b})+\Delta\boldsymbol{h}_{b,u})^{H}\boldsymbol{w}_{b,u}|\leq c_{b,u}|\hat{\boldsymbol{h}}_{b,u}^{H}(\tilde{\boldsymbol{t}}_{b})\boldsymbol{w}_{b,u}|+c_{b,u}|\Delta\boldsymbol{h}_{b,u}^{H}\nonumber\\
&\times\boldsymbol{w}_{b,u}|,
|c_{b,u}(\hat{\boldsymbol{h}}_{b,u}(\tilde{\boldsymbol{t}}_{b})+\Delta\boldsymbol{h}_{b,u})^{H}\boldsymbol{w}_{b,u}|\geq c_{b,u}|\hat{\boldsymbol{h}}_{b,u}^{H}(\tilde{\boldsymbol{t}}_{b})\boldsymbol{w}_{b,u}|-\nonumber\\
&c_{b,u}|\Delta\boldsymbol{h}_{b,u}^{H}\boldsymbol{w}_{b,u}|,\label{proA3}
\end{align}
\end{small}
Thus, based on (\ref{proA3}), we can scale expression (\ref{proA2}) as
\begin{small}
\begin{align}
&\sum\nolimits_{b=1}^{B}(c_{b,u}|\hat{\boldsymbol{h}}_{b,u}^{H}(\tilde{\boldsymbol{t}}_{b})\boldsymbol{w}_{b,u}|-c_{b,u}|\Delta\boldsymbol{h}_{b,u}^{H}\boldsymbol{w}_{b,u}|)^{2}\geq (2^{\gamma_{u}}-1)\left(\sum\nolimits_{b=1}^{B}\right.\nonumber\\
&\left.\left.\sum\nolimits_{u^{\prime}=1,u^{\prime}\neq u}^{U}(c_{b,u^{\prime}}|\hat{\boldsymbol{h}}_{b,u^{\prime}}^{H}(\tilde{\boldsymbol{t}}_{b})\boldsymbol{w}_{b,u^{\prime}}|+c_{b,u^{\prime}}|\Delta\boldsymbol{h}_{b,u^{\prime}}^{H}\boldsymbol{w}_{b,u^{\prime}}|)^{2}\right)+\sigma_{b}^{2}\right).\label{proA4}
\end{align}
\end{small}%
Therefore, the left hand side of (\ref{proA4}) is denoted as
\begin{small}
\begin{align}
&\sum\nolimits_{b=1}^{B}(c_{b,u}|\hat{\boldsymbol{h}}_{b,u}^{H}(\tilde{\boldsymbol{t}}_{b})\boldsymbol{w}_{b,u}|-c_{b,u}|\Delta\boldsymbol{h}_{b,u}^{H}\boldsymbol{w}_{b,u}|)^{2}=\sum\nolimits_{b=1}^{B}(c_{b,u}\times\nonumber\\
&|\hat{\boldsymbol{h}}_{b,u}^{H}(\tilde{\boldsymbol{t}}_{b})\boldsymbol{w}_{b,u}|)^{2}+(c_{b,u}|\Delta\boldsymbol{h}_{b,u}^{H}\boldsymbol{w}_{b,u}|)^{2}-2c_{b,u}^{2}|\hat{\boldsymbol{h}}_{b,u}^{H}(\tilde{\boldsymbol{t}}_{b})\boldsymbol{w}_{b,u}||\Delta\boldsymbol{h}_{b,u}^{H}\boldsymbol{w}_{b,u}|\nonumber\\
&\geq \sum\nolimits_{b=1}^{B}(c_{b,u}|\hat{\boldsymbol{h}}_{b,u}^{H}(\tilde{\boldsymbol{t}}_{b})\boldsymbol{w}_{b,u}|)^{2}-(c_{b,u}|\Delta\boldsymbol{h}_{b,u}^{H}\boldsymbol{w}_{b,u}|)^{2}-2c_{b,u}^{2}\times\nonumber\\
&|\hat{\boldsymbol{h}}_{b,u}^{H}(\tilde{\boldsymbol{t}}_{b})\boldsymbol{w}_{b,u}||\Delta\boldsymbol{h}_{b,u}^{H}\boldsymbol{w}_{b,u}|.\label{proA5}
\end{align}
\end{small}%
Similarly, the right hand side of (\ref{proA4}) is denoted as 
\begin{small}
\begin{align}
&(2^{\gamma_{u}}-1)\left(\sum\nolimits_{b=1}^{B}\right.\sum\nolimits_{u^{\prime}=1,u^{\prime}\neq u}^{U}(c_{b,u^{\prime}}|\hat{\boldsymbol{h}}_{b,u^{\prime}}^{H}(\tilde{\boldsymbol{t}}_{b})\boldsymbol{w}_{b,u^{\prime}}|)^{2}+(c_{b,u}\times\nonumber\\
&\left.\left.|\Delta\boldsymbol{h}_{b,u^{\prime}}^{H}\boldsymbol{w}_{b,u^{\prime}}|)^{2}+2c_{b,u^{\prime}}^{2}|\hat{\boldsymbol{h}}_{b,u^{\prime}}^{H}(\tilde{\boldsymbol{t}}_{b})\boldsymbol{w}_{b,u^{\prime}}||\Delta\boldsymbol{h}_{b,u^{\prime}}^{H}\boldsymbol{w}_{b,u^{\prime}}|\right)+\sigma_{b}^{2}\right).\label{proA6}
\end{align}
\end{small}%
To obtain worst-case communication rate constraint, we further rewritten (\ref{proA4}) as (\ref{proA7}) at the top of this page, then we use the Cauchy-Schwarz inequality to obtain the lower bound of (\ref{proA5}) and the upper bound of (\ref{proA6}), the constraint in (\ref{proA4}) is finally given in  (\ref{proA7}).
\begin{figure*}
\begin{small}
\begin{align}
&\sum\nolimits_{b=1}^{B}(c_{b,u}|\hat{\boldsymbol{h}}_{b,u}^{H}(\tilde{\boldsymbol{t}}_{b})\boldsymbol{w}_{b,u}|)^{2}-(c_{b,u}|\Delta\boldsymbol{h}_{b,u}^{H}\boldsymbol{w}_{b,u}|)^{2}-2c_{b,u}^{2}|\hat{\boldsymbol{h}}_{b,u}^{H}(\tilde{\boldsymbol{t}}_{b})\boldsymbol{w}_{b,u}||\Delta\boldsymbol{h}_{b,u}^{H}\boldsymbol{w}_{b,u}|\geq(2^{\gamma_{u}}-1)\left(\sum\nolimits_{b=1}^{B}\right.\sum\nolimits_{u^{\prime}=1,u^{\prime}\neq u}^{U}(c_{b,u^{\prime}}|\hat{\boldsymbol{h}}_{b,u^{\prime}}^{H}(\tilde{\boldsymbol{t}}_{b})\nonumber\\
&\boldsymbol{w}_{b,u^{\prime}}|)^{2}+\left.\left.(c_{b,u}|\Delta\boldsymbol{h}_{b,u^{\prime}}^{H}\boldsymbol{w}_{b,u^{\prime}}|)^{2}+2c_{b,u^{\prime}}^{2}|\hat{\boldsymbol{h}}_{b,u^{\prime}}^{H}(\tilde{\boldsymbol{t}}_{b})\boldsymbol{w}_{b,u^{\prime}}||\Delta\boldsymbol{h}_{b,u^{\prime}}^{H}\boldsymbol{w}_{b,u^{\prime}}|\right)+\sigma_{b}^{2}\right) \xlongequal{\textrm{Cauchy-Schwarz Inequality} |\Delta\boldsymbol{h}_{b,u}^{H}\boldsymbol{w}_{b,u}|^{2}\leq\|\Delta\boldsymbol{h}_{b,u}\|^{2}\|\boldsymbol{w}_{b,u}\|^{2}}\sum\nolimits_{b=1}^{B}\nonumber\\
&(c_{b,u}|\hat{\boldsymbol{h}}_{b,u}^{H}(\tilde{\boldsymbol{t}}_{b})\boldsymbol{w}_{b,u}|)^{2}-(c_{b,u}\|\Delta\boldsymbol{h}_{b,u}\|^{2}|\boldsymbol{w}_{b,u}|^{2})-2c_{b,u}^{2}|\hat{\boldsymbol{h}}_{b,u}^{H}(\tilde{\boldsymbol{t}}_{b})\boldsymbol{w}_{b,u}|\sqrt{\|\Delta\boldsymbol{h}_{b,u}\|^{2}\|\boldsymbol{w}_{b,u}\|^{2}}\geq (2^{\gamma_{u}}-1)\left(\sum\nolimits_{b=1}^{B}\right.\sum\nolimits_{u^{\prime}=1,u^{\prime}\neq u}^{U}(c_{b,u^{\prime}}|\nonumber\\
&\hat{\boldsymbol{h}}_{b,u^{\prime}}^{H}(\tilde{\boldsymbol{t}}_{b})\boldsymbol{w}_{b,u^{\prime}}|)^{2}+c_{b,u}^{2}\|\Delta\boldsymbol{h}_{b,u^{\prime}}\|^{2}\left.\left.\times\|\boldsymbol{w}_{b,u^{\prime}}\|^{2}+2c_{b,u^{\prime}}^{2}|\hat{\boldsymbol{h}}_{b,u^{\prime}}^{H}(\tilde{\boldsymbol{t}}_{b})\boldsymbol{w}_{b,u^{\prime}}|\sqrt{\|\Delta\boldsymbol{h}_{b,u^{\prime}}\|^{2}\|\boldsymbol{w}_{b,u^{\prime}}\|^{2}}\right)+\sigma_{b}^{2}\right).\label{proA7}
\end{align}    
\end{small}
\hrulefill
\end{figure*}
However, since $\epsilon_{b,u}$ is unknown, we consider how to determine $\epsilon_{b,u}$. Next, we address
$\|\Delta\boldsymbol{h}_{b,u}\|$ in equation (\ref{proA7}), and we have
\begin{figure*}
\begin{small}
\begin{align}
&\hat{\boldsymbol{h}}_{b,u}=\hat{\boldsymbol{A}}_{b,u}(\tilde{\boldsymbol{t}})(\hat{\tilde{\boldsymbol{h}}}_{b,u}+\Delta\tilde{\boldsymbol{h}}_{b,u})=\nonumber\\
&\left[\begin{matrix}
e^{j2\pi/\lambda(x_{1}^{b}\sin(\hat{\theta}_{1}^{b}+\Delta\theta_{1}^{b})\cos(\hat{\phi}_{1}^{b}+\Delta\phi_{1}^{b})+y_{1}^{b}\cos(\hat{\theta}_{1}^{b}+\Delta\theta_{1}^{b}))}&\cdots&e^{j2\pi/\lambda(x_{1}^{b}\sin(\hat{\theta}_{L_{b,u}}^{b}+\Delta\theta_{L_{b,u}}^{b})\cos(\hat{\phi}_{L_{b,u}}^{b}+\Delta\phi_{L_{b,u}}^{b})+y_{1}^{b}\cos(\hat{\theta}_{L_{b,u}}^{b}+\Delta\theta_{L_{b,u}}^{b}))}\\
\vdots&\ddots&\vdots\\
e^{j2\pi/\lambda(x_{N}^{b}\sin(\hat{\theta}_{1}^{b}+\Delta\theta_{1}^{b})\cos(\hat{\phi}_{1}^{b}+\Delta\phi_{1}^{b})+y_{N}^{b}\cos(\hat{\theta}_{1}^{b}+\Delta\theta_{1}^{b}))}&\cdots&e^{j2\pi/\lambda(x_{N}^{b}\sin(\hat{\theta}_{L_{b,u}}^{b}+\Delta\theta_{L_{b,u}}^{b})\cos(\hat{\phi}_{L_{b,u}}^{b}+\Delta\phi_{L_{b,u}}^{b})+y_{N}^{b}\cos(\hat{\theta}_{L_{b,u}}^{b}+\Delta\theta_{L_{b,u}}^{b}))}\\
\end{matrix}\right]\nonumber\\
&\times\left[\hat{\tilde{h}}_{b,u,1}+\Delta\tilde{h}_{b,u,1},\cdots,\hat{\tilde{h}}_{b,u,L_{b,u}}+\Delta\tilde{h}_{b,u,L_{b,u}}\right]^{T}.\label{proA8}
\end{align}    
\end{small}
\hrulefill
\end{figure*}
Then, $\|\Delta\boldsymbol{h}_{b,u}\|^{2}$ is rewritten as
\begin{small}
\begin{align}
&|\Delta\boldsymbol{h}_{b,u}|^{2}=|\hat{\boldsymbol{h}}_{b,u}(\tilde{\boldsymbol{t}}_{b})-\boldsymbol{h}_{b,u}(\tilde{\boldsymbol{t}}_{b})|^{2}=\sum\nolimits_{n=1}^{N}|\sum\nolimits_{i=1}^{L_{b,u}}\nonumber\\
&e^{j2\pi/\lambda(x_{t,b}^{n}\sin(\hat{\theta}_{i}^{b}+\Delta\theta_{i}^{b})\cos(\hat{\phi}_{i}^{b}+\Delta\phi_{i}^{b})+y_{t,b}^{n}\cos(\hat{\theta}_{i}^{b}+\Delta\theta_{i}^{b})\sin(\hat{\phi}_{i}^{b}+\Delta\phi_{i}^{b}))}\nonumber\\
&(\hat{\tilde{h}}_{b,u,i}+\Delta\tilde{h}_{b,u,i})-e^{j2\pi/\lambda(x_{j}^{b}\sin(\hat{\theta}_{i}^{b})\cos(\hat{\phi}_{i}^{b})+y_{j}^{b}\cos(\hat{\theta}_{i}^{b})\sin(\hat{\phi}_{i}^{b}))}\hat{\tilde{h}}_{b,u,i}|^{2}.\label{proA9}
\end{align}
\end{small}%
Then, we can use the first order Taylor series of 
$\sin(\hat{\theta}_{i}^{b} +\Delta\theta_{i}^{b})=\sin(\hat{\theta}_{i}^{b})+\Delta\theta_{i}^{b}\cos(\hat{\theta}_{i}^{b})$ and 
$\cos(\hat{\theta}_{i}^{b} + \Delta\theta_{i}^{b}) = \cos(\hat{\theta}_{i}^{b}) - \Delta\theta_{i}^{b} \sin(\hat{\theta}_{i}^{b})$ to obtain
\begin{small}
\begin{align}
&\sum_{j=1}^{N}|\sum\nolimits_{i=1}^{L_{b,u}}e^{j2\pi/\lambda(x_{j}^{b}(-\Delta\phi_{i}^{b}\sin(\hat{\phi}_{i}^{b})\sin(\hat{\theta}_{t}^{b})+\Delta\theta_{i}^{b}\cos(\hat{\phi}_{i}^{b})\cos(\hat{\theta}_{t}^{b}))}\nonumber\\
&e^{j2\pi/\lambda-x_{j}^{b}(\Delta\phi_{i}^{b}\Delta\theta_{i}^{b}\sin(\hat{\phi}_{i}^{b})\cos(\hat{\theta}_{i}^{b}))-y_{j}^{b}\Delta\theta_{i}^{b}\sin(\hat{\theta}_{i}^{b})))}(\hat{\tilde{h}}_{b,u,i}+\Delta\tilde{h}_{b,u,i})\nonumber\\
&+e^{j2\pi/\lambda(x_{j}^{b}\sin(\hat{\theta}_{i}^{b})\cos(\hat{\phi}_{i}^{b})+y_{j}^{b}\cos(\hat{\theta}_{i}^{b}))}\Delta\tilde{h}_{b,u,i}|^{2}.\label{proA10}
\end{align}
\end{small}
Then, by applying the triangle inequality, we can further enlarge the result to obtain the upper bound of $|\Delta\boldsymbol{h}_{b,u}|^{2}$, and the upper bound is given in (\ref{proA11}) at the top of next page.
\begin{figure*}
\begin{small}
\begin{align}
&\sum\nolimits_{j=1}^{N}|\sum\nolimits_{i=1}^{L_{b,u}}e^{j2\pi/\lambda(x_{j}^{b}(-\Delta\phi_{i}^{b}\sin(\hat{\phi}_{i}^{b})\sin(\hat{\theta}_{t}^{b})+\Delta\theta_{i}^{b}\cos(\hat{\phi}_{i}^{b})\cos(\hat{\theta}_{t}^{b}))}e^{j2\pi/\lambda-x_{j}^{b}(\Delta\phi_{i}^{b}\Delta\theta_{i}^{b}\sin(\hat{\phi}_{i}^{b})\cos(\hat{\theta}_{i}^{b}))-y_{j}^{b}\Delta\theta_{i}^{b}\sin(\hat{\theta}_{i}^{b})))}(\hat{\tilde{h}}_{b,u,i}+\Delta\tilde{h}_{b,u,i})|^{2}\nonumber\\
&+|\sum\nolimits_{i=1}^{L_{b,u}}e^{j2\pi/\lambda(x_{j}^{b}\sin(\hat{\theta}_{i}^{b})\cos(\hat{\phi}_{i}^{b})+y_{j}^{b}\cos(\hat{\theta}_{i}^{b}))}\Delta\tilde{h}_{b,u,i}|^{2}\leq\sum\nolimits_{j=1}^{N}|\sum\nolimits_{i=1}^{L_{b,u}}e^{j2\pi/\lambda(x_{j}^{b}(-\Delta\phi_{i}^{b}\sin(\hat{\phi}_{i}^{b})\sin(\hat{\theta}_{t}^{b})+\Delta\theta_{i}^{b}\cos(\hat{\phi}_{i}^{b})\cos(\hat{\theta}_{t}^{b}))}\nonumber\\
&e^{j2\pi/\lambda-x_{j}^{b}(\Delta\phi_{i}^{b}\Delta\theta_{i}^{b}\sin(\hat{\phi}_{i}^{b})\cos(\hat{\theta}_{i}^{b}))-y_{j}^{b}\Delta\theta_{i}^{b}\sin(\hat{\theta}_{i}^{b})))}\hat{\tilde{h}}_{b,u,i}|^{2}+2NL_{b,u}\epsilon_{b,u}=N\|\hat{\tilde{\boldsymbol{h}}}_{b,u,i}\|^{2}+2NL_{b,u}\bar{\epsilon}_{b,u}.\label{proA11}
\end{align}     
\end{small}
\hrulefill
\end{figure*}
Therefore, the upper bound of $|\Delta\boldsymbol{h}_{b,u}|^{2}$ is expressed as
\begin{small}
\begin{align}
|\Delta\boldsymbol{h}_{b,u}|^{2}\leq N\|\hat{\tilde{\boldsymbol{h}}}_{b,u}\|^{2}+2NL_{b,u}\bar{\epsilon}_{b,u}.\label{proA12} 
\end{align}
\end{small}%
Finally, by substituting the results from (\ref{proA12}) into equation (\ref{proA7}), we obtain the worst-case communication rate constraint, and it is given in (\ref{pro13}). The \textbf{Theorem}~\ref{th1} is proofed in \textbf{Appendix}~\ref{APP1}

\section{The proof of (\ref{pro21})}\label{appB}

%Based on (10), each sub-matrix of $\boldsymbol{F}_{n}(\boldsymbol{q}_{n})$ is represented as $\boldsymbol{F}^{n}_{\boldsymbol{p}\boldsymbol{p}}=\boldsymbol{F}^{n,D}_{\boldsymbol{p}\boldsymbol{p}}+\boldsymbol{F}^{n,B}_{\boldsymbol{p}\boldsymbol{p}}$, $\boldsymbol{F}^{n}_{\boldsymbol{p}\Delta\boldsymbol{u}_{n}}=\boldsymbol{F}^{n,D}_{\boldsymbol{p}\Delta\boldsymbol{u}_{n}}+\boldsymbol{F}^{n,B}_{\boldsymbol{p}\Delta\boldsymbol{u}_{n}}$ and $\boldsymbol{F}^{n}_{\Delta\boldsymbol{u}_{n}\Delta\boldsymbol{u}_{n}}=\boldsymbol{F}^{n,D}_{\Delta\boldsymbol{u}_{n}\Delta\boldsymbol{u}_{n}}+\boldsymbol{F}^{n,B}_{\Delta\boldsymbol{u}_{n}\Delta\boldsymbol{u}_{n}}$. $\boldsymbol{F}^{n,D}_{\boldsymbol{p}\boldsymbol{p}}$ and $\boldsymbol{F}^{n,D}_{\boldsymbol{p}\Delta\boldsymbol{u}_{n}}$, and $\boldsymbol{F}^{n,D}_{\Delta\boldsymbol{u}_{n}\Delta\boldsymbol{u}_{n}}$ are sub-matrices of $\boldsymbol{F}^{n,D}_{\boldsymbol{p}\boldsymbol{p}}$. $\boldsymbol{F}^{n,B}_{\boldsymbol{p}\boldsymbol{p}}$ and $\boldsymbol{F}^{n,B}_{\boldsymbol{p}\Delta\boldsymbol{u}_{n}}$, and $\boldsymbol{F}^{n,B}_{\Delta\boldsymbol{u}_{n}\Delta\boldsymbol{u}_{n}}$ are sub-matrices of $\boldsymbol{F}^{n,B}_{\boldsymbol{p}\boldsymbol{p}}$. Next, the submatrices of $\boldsymbol{F}_{n}^{D}(\boldsymbol{q}_{n})$ and $\boldsymbol{F}_{n}^{B}(\boldsymbol{q}_{n})$ are derived, respectively.
%

As stated in \cite{b30}, the $(o_{1}, o_{2})$-th entry of the OFIM and PFIM is expressed as, respectively
\begin{small}
\begin{align}
&\boldsymbol{\Xi}^{o}_{b}[o_{1}, o_{2}]=-\mathbb{E}\left(\frac{\partial^{2}\ln p(\tilde{\boldsymbol{y}}_{b};\boldsymbol{\zeta}_{b})}{\partial\boldsymbol{\zeta}_{b}[o_{1}]\partial\boldsymbol{\zeta}_{b}[o_{2}]}\right)\nonumber\\
&=\frac{2}{\sigma_{b}^{2}}\mathrm{Re}\left\{\frac{\partial\tilde{\boldsymbol{x}}_{b}^{H}}{\partial\boldsymbol{\zeta}_{b}[o_{1}]}\frac{\partial\tilde{\boldsymbol{x}}_{b}}{\partial\boldsymbol{\zeta}_{b}[o_{2}]}\right\}+\underbrace{\mathrm{Tr}\left(\boldsymbol{I}_{b}^{-1}\frac{\partial\boldsymbol{I}_{b}}{\partial\boldsymbol{\zeta}_{b}[o_{1}]}\boldsymbol{I}_{b}^{-1}\frac{\partial\boldsymbol{I}_{b}}{\partial\boldsymbol{\zeta}_{b}[o_{2}]}\right)}_{\boldsymbol{0}}\nonumber\\
&=\frac{2}{\sigma_{b}^{2}}\mathrm{Re}\left\{\frac{\partial\tilde{\boldsymbol{x}}_{b}^{H}}{\partial\boldsymbol{\zeta}_{b}[o_{1}]}\frac{\partial\tilde{\boldsymbol{x}}_{b}}{\partial\boldsymbol{\zeta}_{b}[o_{2}]}\right\}\label{proB1}
\end{align} 
\end{small}
and
\begin{small}
\begin{align}
&\boldsymbol{\Xi}^{p}_{b}[o_{1}, o_{2}]=-\mathbb{E}\left(\frac{\partial^{2}\ln p(\Delta\boldsymbol{\xi}_{b})}{\partial\boldsymbol{\zeta}_{b}[o_{1}]\partial\boldsymbol{\zeta}_{b}[o_{2}]}\right)\nonumber\\
&=\frac{2}{\sigma_{\xi}^{2}}\mathrm{Re}\left\{\frac{\partial\Delta\boldsymbol{\xi}_{b}^{H}}{\partial\boldsymbol{\zeta}_{b}[o_{1}]}\frac{\partial\Delta\boldsymbol{\xi}_{b}}{\partial\boldsymbol{\zeta}_{b}[o_{2}]}\right\}+\underbrace{\mathrm{Tr}\left(\boldsymbol{I}_{b}^{-1}\frac{\partial\boldsymbol{I}_{b}}{\partial\boldsymbol{\zeta}_{b}[o_{1}]}\boldsymbol{I}_{b}^{-1}\frac{\partial\boldsymbol{I}_{b}}{\partial\boldsymbol{\zeta}_{b}[o_{2}]}\right)}_{\boldsymbol{0}}\nonumber\\
&=\frac{2}{\sigma_{\xi}^{2}}\mathrm{Re}\left\{\frac{\partial\Delta\boldsymbol{\xi}_{b}^{H}}{\partial\boldsymbol{\zeta}_{b}[o_{1}]}\frac{\partial\Delta\boldsymbol{\xi}_{b}}{\partial\boldsymbol{\zeta}_{b}[o_{2}]}\right\},\label{proB2}
\end{align}    
\end{small}%
in which $\boldsymbol{\zeta}_{b}[o_{1}]$ and $\boldsymbol{\zeta}_{b}[o_{1}]$ are the $o_{1}$-th element and $o_{2}$-th element of $\boldsymbol{\zeta}_{b}$, respectively. In (\ref{proB1}) and (\ref{proB2}), we observe that to determine the expressions of (\ref{proB1}) and (\ref{proB2}), it is necessary to compute the derivatives of $\tilde{\boldsymbol{x}}_{b}$ and $\Delta\boldsymbol{\xi}_{b}$ to the unknown parameter $\boldsymbol{\zeta}_{b}$. Specifically, based on (\ref{pro15}), the derivatives of $\tilde{\boldsymbol{x}}_{b}$ and $\Delta\boldsymbol{\xi}_{b}$ with respect to $x_{T}$ are expressed as {follows:}
\begin{small}
\begin{align}
&\frac{\partial \tilde{\boldsymbol{x}}_{b}}{\partial x_{T}}=\sum\nolimits_{b^{\prime}=1}^{B}\frac{\partial \mathrm{vec}(\boldsymbol{H}_{b,b^{\prime}}(\tilde{\boldsymbol{r}}_{b^{\prime}},\tilde{\boldsymbol{t}}_{b})\boldsymbol{\Omega}_{b})}{\partial x_{T}}+\frac{\partial\hat{\xi}_{b,b^{\prime}}^{p}}{\partial x_{T}}\times\nonumber\\
&\mathrm{vec}(\boldsymbol{H}_{b,b^{\prime}}(\tilde{\boldsymbol{r}}_{b^{\prime}},\tilde{\boldsymbol{t}}_{b})\bar{\boldsymbol{\Omega}}_{b}),~\frac{\partial \tilde{\boldsymbol{\xi}}_{b}}{\partial x_{T}}=\boldsymbol{0}, \label{proB3}
\end{align}    
\end{small}
in which $\boldsymbol{\Omega}_{b}=[\bar{c}_{b}\boldsymbol{W}_{b}e^{-jf_{1}\hat{\xi}_{b,b^{\prime}}^{p}}\boldsymbol{s}_{b}(f_{1}),\cdots,\bar{c}_{b}\boldsymbol{W}_{b}e^{-jf_{\bar{S}}\hat{\xi}_{b,b^{\prime}}^{p}}\\
\boldsymbol{s}_{b}(f_{\bar{S}})]$ and
$\bar{\boldsymbol{\Omega}}_{b}=[-jf_{1}\bar{c}_{b}\boldsymbol{W}_{b}e^{-jf_{1}\hat{\xi}_{b,b^{\prime}}^{p}}\boldsymbol{s}_{b}(f_{1}),\cdots,-jf_{M}\bar{c}_{b}\\
\boldsymbol{W}_{b}e^{-jf_{\bar{S}}\hat{\xi}_{b,b^{\prime}}^{p}}\boldsymbol{s}_{b}(f_{\bar{S}})]$. According to vectorization rules $\frac{\partial \mathrm{vec}(\boldsymbol{A}\boldsymbol{B})}{\partial x}= \mathrm{vec}(\frac{\partial\boldsymbol{A}}{\partial x}\boldsymbol{B})$ in\cite{b37}. Thus, we can determine the expression for A after computing the derivative of  $\boldsymbol{H}_{b,b^{\prime}}(\tilde{\boldsymbol{r}}_{b^{\prime}},\tilde{\boldsymbol{t}}_{b})$ and $\tau_{b,b^{\prime}}$ with respect to $x_{T}${,~i.e.,}
\begin{small}
\begin{align}
&\frac{\partial \boldsymbol{H}(\tilde{\boldsymbol{r}}_{b^{\prime}},\tilde{\boldsymbol{t}}_{b})}{\partial x_{T}}=\alpha_{b,b^{\prime}}\frac{\partial \boldsymbol{a}(\tilde{\boldsymbol{r}}_{b^{\prime}})}{\partial x_{T}}\boldsymbol{a}(\tilde{\boldsymbol{t}}_{b})^{H}+\alpha_{b,b^{\prime}}\boldsymbol{a}(\tilde{\boldsymbol{r}}_{b^{\prime}})\frac{\partial \boldsymbol{a}(\tilde{\boldsymbol{t}}_{b})}{\partial x_{T}}^{H}\\
&\frac{\partial \tau_{b,b^{\prime}}}{\partial x_{T}}=1/c(x_{T}-x_{b})/v_{b}+(x_{T}-x_{b^{\prime}})/v_{b^{\prime}}). \label{proB4}
\end{align}
\end{small}%
According to the chain rule of differentiation\cite{b37}, $\frac{\partial \boldsymbol{a}(\tilde{\boldsymbol{r}}_{b^{\prime}})}{\partial x_{T}}$ and $\frac{\partial \boldsymbol{a}(\tilde{\boldsymbol{t}}_{b})}{\partial x_{T}}$ are given by
\begin{small}
\begin{align}
&\frac{\partial \boldsymbol{a}(\tilde{\boldsymbol{r}}_{b^{\prime}})}{\partial x_{T}}=\left(\frac{\partial\rho^{b^{\prime}}(\boldsymbol{r}_{b^{\prime}}^{m})}{\partial \bar{\theta}_{b}}\frac{\partial\bar{\theta}_{b}}{x_{T}}+\frac{\partial\rho^{b^{\prime}}(\boldsymbol{r}_{b^{\prime}}^{m})}{\partial \bar{\phi}_{b}}\frac{\partial\bar{\phi}_{b}}{x_{T}}\right)\bar{\boldsymbol{a}}(\tilde{\boldsymbol{r}}_{b^{\prime}})\odot\boldsymbol{a}(\tilde{\boldsymbol{r}}_{b^{\prime}})\nonumber\\
&\frac{\partial \boldsymbol{a}(\tilde{\boldsymbol{t}}_{b})}{\partial x_{T}}=\left(\frac{\partial\rho^{b}(\boldsymbol{t}_{b}^{n})}{\partial \theta_{b}}\frac{\partial\theta_{b}}{x_{T}}+\frac{\partial\rho^{b}(\boldsymbol{t}_{b}^{n})}{\partial \phi_{b}}\frac{\partial\phi_{b}}{x_{T}}\right)\bar{\boldsymbol{a}}(\tilde{\boldsymbol{t}}_{b})\odot\boldsymbol{a}(\tilde{\boldsymbol{t}}_{b}), \label{proB5}
\end{align}
\end{small}%
where $\bar{\boldsymbol{a}}(\tilde{\boldsymbol{r}}_{b^{\prime}})=[j2\pi/\lambda\rho_{1}(\boldsymbol{t}),\cdots,j2\pi/\lambda\rho_{N}(\boldsymbol{t})]$ and
$\bar{\boldsymbol{a}}(\tilde{\boldsymbol{r}}_{b^{\prime}})=[j2\pi/\lambda\rho_{1}(\boldsymbol{t}),\cdots,j2\pi/\lambda\rho_{N}(\boldsymbol{t})]$. Moreover, $\frac{\partial \rho^{b}(\boldsymbol{t}_{b}^{n})}{\partial \theta_{b}}$, $\frac{\partial \rho^{b}(\boldsymbol{t}_{b}^{n})}{\partial \phi_{b}}$, $\frac{\partial \rho^{b^{\prime}}(\boldsymbol{r}_{b^{\prime}}^{m})}{\partial \bar{\theta}_{b}}$ and $\frac{\partial \rho^{b^{\prime}}(\boldsymbol{r}_{b^{\prime}}^{m})}{\partial \bar{\phi}_{b}}$ are expressed as
\begin{small}
\begin{align}
&\frac{\partial \rho^{b}(\boldsymbol{t}_{b}^{n})}{\partial \theta_{b}}=x_{n}^{b}\cos\theta^{b}\cos\phi^{b}-y_{n}^{b}\sin\theta^{b},\nonumber\\
&\frac{\partial \rho^{b^{\prime}}(\boldsymbol{r}_{b^{\prime}}^{m})}{\partial \bar{\theta}_{b}}=x_{m}^{b}\cos\theta^{b}\cos\phi^{b}-y_{m}^{b}\sin\theta^{b},\nonumber\\
&\frac{\partial \rho^{b}(\boldsymbol{t}_{b}^{n})}{\partial \phi_{b}}=-x_{n}^{b}\sin\theta^{b}\sin\phi^{b}, \frac{\partial \rho^{b^{\prime}}(\boldsymbol{r}_{b^{\prime}}^{m})}{\partial \bar{\phi}_{b}}=-x_{m}^{b}\sin\theta^{b}\sin\phi^{b}, \label{proB6}
\end{align}
\end{small}%
$\frac{\partial \theta_{b}}{\partial x_{T}}$, $\frac{\partial \phi_{b}}{\partial x_{T}}$, $\frac{\partial \bar{\theta}_{b^{\prime}}}{\partial x_{T}}$ and $\frac{\partial \bar{\phi}_{b^{\prime}}}{\partial x_{T}}$ are given at the top of this page.
\begin{figure*}
\begin{small}
\begin{align}
&\frac{\partial \theta_{b}}{\partial x_{T}}=\frac{y_b - x_b}{(x_T - y_b)^2 + (x_T - x_b)^2}
,~\frac{\partial \phi_{b}}{\partial x_{T}}= -\frac{z_b \cdot \left( (x_T - x_b) + (x_T - y_b) \right)}{(x_T - x_b)^2 + (x_T - y_b)^2 + z_b^2 \cdot ((x_T - x_b)^2 + (x_T - y_b)^2)}
,\nonumber\\
&\frac{\partial \bar{\theta}_{b^{\prime}}}{\partial x_{T}}=\frac{y_b^{\prime} - x_b^{\prime}}{(x_T - y_b^{\prime})^2 + (x_T - x_b^{\prime})^2}
,~\frac{\partial \bar{\phi}_{b^{\prime}}}{\partial x_{T}}= -\frac{z_b^{\prime} \cdot \left( (x_T - x_b^{\prime}) + (x_T - y_b^{\prime}) \right)}{(x_T - x_b^{\prime})^2 + (x_T - y_b^{\prime})^2 + (z_b^{\prime})^2 \cdot ((x_T - x_b^{\prime})^2 + (x_T - y_b^{\prime})^2)},\label{proB7}
\end{align}
\end{small} 
\hrulefill
\end{figure*}
Similarly, we can determine the derivative of  $\boldsymbol{H}_{b,b^{\prime}}(\tilde{\boldsymbol{r}}_{b^{\prime}},\tilde{\boldsymbol{t}}_{b})$ and $\tau_{b,b^{\prime}}$ with respect to $y_{T}$.
The derivative of $\tilde{\boldsymbol{x}}_{b}$ with respect to $\Delta\boldsymbol{\xi}_{b}[o_{3}]$ is expressed as
\begin{small}
\begin{align}
\frac{\partial \tilde{\boldsymbol{x}}_{b}}{\partial \Delta\boldsymbol{\xi}_{b}[o_{3}]}=\mathrm{vec}(\boldsymbol{H}_{b,b^{\prime}}\bar{\boldsymbol{\Omega}}_{b}), \label{proB8}
\end{align}    
\end{small}%
in which $\Delta\boldsymbol{\xi}_{b}[o_{3}]$ is the $o_{3}$-th element of $\Delta\boldsymbol{\xi}_{b}$. Then, $\frac{\partial^{2}p(\tilde{\boldsymbol{y}}_{b};\boldsymbol{\zeta}_{b})}{\partial\boldsymbol{\zeta}_{b}[o_{1}]\partial\boldsymbol{\zeta}_{b}[o_{2}]}$ can be further specifically expressed as
\begin{small}
\begin{align}
\frac{\partial^{2}p(\tilde{\boldsymbol{y}}_{b};\boldsymbol{\zeta}_{b})}{\partial\boldsymbol{\zeta}_{b}[o_{1}]\partial\boldsymbol{\zeta}_{b}[o_{2}]}=\left[\begin{matrix}
\frac{\partial^{2}p(\tilde{\boldsymbol{y}}_{b};\boldsymbol{\zeta}_{b})}{\partial x_{T}\partial x_{T}}&\frac{\partial^{2}p(\tilde{\boldsymbol{y}}_{b};\boldsymbol{\zeta}_{b})}{\partial x_{T}\partial y_{T}}&\frac{\partial^{2}p(\tilde{\boldsymbol{y}}_{b};\boldsymbol{\zeta}_{b})}{\partial x_{T}\partial\Delta\boldsymbol{\xi}_{b}}\\
\frac{\partial^{2}p(\tilde{\boldsymbol{y}}_{b};\boldsymbol{\zeta}_{b})}{\partial y_{T}\partial x_{T}}&\frac{\partial^{2}p(\tilde{\boldsymbol{y}}_{b};\boldsymbol{\zeta}_{b})}{\partial y_{T}\partial y_{T}}&\frac{\partial^{2}p(\tilde{\boldsymbol{y}}_{b};\boldsymbol{\zeta}_{b})}{\partial y_{T}\partial\Delta\boldsymbol{\xi}_{b}}\\
\frac{\partial^{2}p(\tilde{\boldsymbol{y}}_{b};\boldsymbol{\zeta}_{b})}{\partial x_{T}\partial\Delta\boldsymbol{\xi}_{b}}&\frac{\partial^{2}p(\tilde{\boldsymbol{y}}_{b};\boldsymbol{\zeta}_{b})}{\partial y_{T}\partial\Delta\boldsymbol{\xi}_{b}}&\frac{\partial^{2}p(\tilde{\boldsymbol{y}}_{b};\boldsymbol{\zeta}_{b})}{\partial \Delta\boldsymbol{\xi}_{b}\partial\Delta\boldsymbol{\xi}_{b}}
\end{matrix}\right]. \label{proB9}
\end{align}    
\end{small}%
Based on (\ref{proB1}) and (\ref{proB2}), 
$\frac{\partial^{2}p(\tilde{\boldsymbol{y}}_{b};\boldsymbol{\zeta}_{b})}{\partial x_{T}\partial x_{T}}$ is computed as (\ref{proB10}) at the top of next page.
\begin{figure*}
\begin{small}
\begin{align}
&\frac{\partial^{2}p(\tilde{\boldsymbol{y}}_{b};\boldsymbol{\zeta}_{b})}{\partial x_{T}\partial x_{T}}=\frac{2}{\sigma_{b}^{2}}\mathrm{Re}\left\{\left(\sum\nolimits_{b^{\prime}=1}^{B}\frac{\partial \mathrm{vec}(\boldsymbol{H}_{b,b^{\prime}}(\tilde{\boldsymbol{r}}_{b^{\prime}},\tilde{\boldsymbol{t}}_{b})\boldsymbol{\Omega}_{b})}{\partial x_{T}}+\frac{\partial\hat{\xi}_{b,b^{\prime}}^{p}}{\partial x_{T}}\mathrm{vec}(\boldsymbol{H}_{b,b^{\prime}}(\tilde{\boldsymbol{r}}_{b^{\prime}},\tilde{\boldsymbol{t}}_{b})\bar{\boldsymbol{\Omega}}_{b})\right)^{H}\left(\sum\nolimits_{b^{\prime}=1}^{B}\frac{\partial \mathrm{vec}(\boldsymbol{H}_{b,b^{\prime}}(\tilde{\boldsymbol{r}}_{b^{\prime}},\tilde{\boldsymbol{t}}_{b})\boldsymbol{\Omega}_{b})}{\partial x_{T}}+\right.\right.\nonumber\\
&\left.\left.\frac{\partial\hat{\xi}_{b,b^{\prime}}^{p}}{\partial x_{T}}\mathrm{vec}(\boldsymbol{H}_{b,b^{\prime}}(\tilde{\boldsymbol{r}}_{b^{\prime}},\tilde{\boldsymbol{t}}_{b})\bar{\boldsymbol{\Omega}}_{b})\right)\right\}=\frac{2}{\sigma_{b}^{2}}\mathrm{Re}\left\{\sum\nolimits_{m=1}^{M}\mathrm{Tr}\left(\sum\nolimits_{s=1}^{\bar{S}}|\boldsymbol{s}_{b}(f_{s})|^{2}\left(\frac{\partial \boldsymbol{H}(\tilde{\boldsymbol{r}}_{b^{\prime}},\tilde{\boldsymbol{t}}_{b})}{\partial x_{T}}-jf_{s}\frac{\partial \tau_{b,b^{\prime}}}{\partial x_{T}}\boldsymbol{H}(\tilde{\boldsymbol{r}}_{b^{\prime}},\tilde{\boldsymbol{t}}_{b})\right)^{H}\left(\frac{\partial \boldsymbol{H}(\tilde{\boldsymbol{r}}_{b^{\prime}},\tilde{\boldsymbol{t}}_{b})}{\partial x_{T}}-\right.\right.\right.\nonumber\\
&\left.\left.jf_{s}\frac{\partial \tau_{b,b^{\prime}}}{\partial x_{T}}\boldsymbol{H}(\tilde{\boldsymbol{r}}_{b^{\prime}},\tilde{\boldsymbol{t}}_{b}))\boldsymbol{w}\boldsymbol{w}^{H}\right)\right\}. \label{proB10}
\end{align}    
\end{small}
\hrulefill
\end{figure*}
Similarly, $\frac{\partial^{2}p(\tilde{\boldsymbol{y}}_{b};\boldsymbol{\zeta}_{b})}{\partial x_{T}\partial x_{T}}$ and $\frac{\partial^{2}p(\tilde{\boldsymbol{y}}_{b};\boldsymbol{\zeta}_{b})}{\partial x_{T}\partial x_{T}}$ are denoted as (\ref{proB11}) at the top of next page.
\begin{figure*}
\begin{small}
\begin{align}
&\frac{\partial^{2}p(\tilde{\boldsymbol{y}}_{b};\boldsymbol{\zeta}_{b})}{\partial x_{T}\partial y_{T}}=\frac{2}{\sigma_{b}^{2}}\mathrm{Re}\left\{\sum_{m=1}^{M}\mathrm{Tr}\left(\sum_{s=1}^{\bar{S}}|\boldsymbol{s}_{b}(f_{s})|^{2}\left(\frac{\partial \boldsymbol{H}(\tilde{\boldsymbol{r}}_{b^{\prime}},\tilde{\boldsymbol{t}}_{b})}{\partial x_{T}}-jf_{s}\frac{\partial \tau_{b,b^{\prime}}}{\partial x_{T}}\boldsymbol{H}(\tilde{\boldsymbol{r}}_{b^{\prime}},\tilde{\boldsymbol{t}}_{b})\right)^{H}\left(\frac{\partial \boldsymbol{H}(\tilde{\boldsymbol{r}}_{b^{\prime}},\tilde{\boldsymbol{t}}_{b})}{\partial y_{T}}-jf_{s}\frac{\partial \tau_{b,b^{\prime}}}{\partial y_{T}}\boldsymbol{H}(\tilde{\boldsymbol{r}}_{b^{\prime}},\tilde{\boldsymbol{t}}_{b})\right)\boldsymbol{w}\boldsymbol{w}^{H}\right)\right\},\nonumber\\
&\frac{\partial^{2}p(\tilde{\boldsymbol{y}}_{b};\boldsymbol{\zeta}_{b})}{\partial y_{T}\partial x_{T}}=\frac{2}{\sigma_{b}^{2}}\mathrm{Re}\left\{\sum_{m=1}^{M}\mathrm{Tr}\left(\sum_{s=1}^{\bar{S}}|\boldsymbol{s}_{b}(f_{s})|^{2}\left(\frac{\partial \boldsymbol{H}(\tilde{\boldsymbol{r}}_{b^{\prime}},\tilde{\boldsymbol{t}}_{b})}{\partial y_{T}}-jf_{s}\frac{\partial \tau_{b,b^{\prime}}}{\partial y_{T}}\boldsymbol{H}(\tilde{\boldsymbol{r}}_{b^{\prime}},\tilde{\boldsymbol{t}}_{b})\right)^{H}\left(\frac{\partial \boldsymbol{H}(\tilde{\boldsymbol{r}}_{b^{\prime}},\tilde{\boldsymbol{t}}_{b})}{\partial x_{T}}-jf_{s}\frac{\partial \tau_{b,b^{\prime}}}{\partial x_{T}}\boldsymbol{H}(\tilde{\boldsymbol{r}}_{b^{\prime}},\tilde{\boldsymbol{t}}_{b})\right)\boldsymbol{w}\boldsymbol{w}^{H}\right)\right\}. \label{proB11}
\end{align}      
\end{small}
\hrulefill
\end{figure*}
Similarly, according to (\ref{proB10}) and (\ref{proB11}), $\frac{\partial^{2}p(\tilde{\boldsymbol{y}}_{b};\boldsymbol{\zeta}_{b})}{\partial x_{T}\partial\Delta\boldsymbol{\xi}_{b}}$ and $\frac{\partial^{2}p(\tilde{\boldsymbol{y}}_{b};\boldsymbol{\zeta}_{b})}{\partial x_{T}\partial\Delta\boldsymbol{\xi}_{b}}$ are given by (\ref{proB12}) at the top of next page.
\begin{figure*}
\begin{small}
\begin{align}
&\frac{\partial^{2}p(\tilde{\boldsymbol{y}}_{b};\boldsymbol{\zeta}_{b})}{\partial x_{T}\partial\Delta\boldsymbol{\xi}_{b}}=\frac{2}{\sigma_{b}^{2}}\mathrm{Re}\left\{\sum_{m=1}^{M}\mathrm{Tr}\left(\sum\nolimits_{s=1}^{\bar{S}}|\boldsymbol{s}_{b}(f_{s})|^{2}\left(\frac{\partial \boldsymbol{H}(\tilde{\boldsymbol{r}}_{b^{\prime}},\tilde{\boldsymbol{t}}_{b})}{\partial x_{T}}-jf_{s}\frac{\partial \tau_{b,b^{\prime}}}{\partial x_{T}}\boldsymbol{H}(\tilde{\boldsymbol{r}}_{b^{\prime}},\tilde{\boldsymbol{t}}_{b})\right)^{H}\boldsymbol{H}(\tilde{\boldsymbol{r}}_{b^{\prime}},\tilde{\boldsymbol{t}}_{b})\boldsymbol{w}\boldsymbol{w}^{H}\right)\right\},\nonumber\\
&\frac{\partial^{2}p(\tilde{\boldsymbol{y}}_{b};\boldsymbol{\zeta}_{b})}{\partial y_{T}\partial\Delta\boldsymbol{\xi}_{b}}=\frac{2}{\sigma_{b}^{2}}\mathrm{Re}\left\{\sum_{m=1}^{M}\mathrm{Tr}\left(\sum\nolimits_{s=1}^{\bar{S}}|\boldsymbol{s}_{b}(f_{s})|^{2}\left(\frac{\partial \boldsymbol{H}(\tilde{\boldsymbol{r}}_{b^{\prime}},\tilde{\boldsymbol{t}}_{b})}{\partial y_{T}}-jf_{s}\frac{\partial \tau_{b,b^{\prime}}}{\partial y_{T}}\boldsymbol{H}(\tilde{\boldsymbol{r}}_{b^{\prime}},\tilde{\boldsymbol{t}}_{b})\right)^{H}\boldsymbol{H}(\tilde{\boldsymbol{r}}_{b^{\prime}},\tilde{\boldsymbol{t}}_{b})\boldsymbol{w}\boldsymbol{w}^{H}\right)\right\}. \label{proB12}
\end{align}   
\end{small}
\hrulefill
\end{figure*}
Then, when $o_{3}\neq o_{3}^{\prime}$, $\frac{\partial^{2}p(\tilde{\boldsymbol{y}}_{b};\boldsymbol{\zeta}_{b})}{\partial \Delta\boldsymbol{\xi}_{b}[o_{3}]\partial\Delta\boldsymbol{\xi}_{b}[o_{3}]}=0$, when $o_{3}= o_{3}^{\prime}$, $\frac{\partial^{2}p(\tilde{\boldsymbol{y}}_{b};\boldsymbol{\zeta}_{b})}{\partial \Delta\boldsymbol{\xi}_{b}[o_{3}]\partial\Delta\boldsymbol{\xi}_{b}[o_{3}]}$ is expressed as
\begin{small}
\begin{align}
&\frac{\partial^{2}p(\tilde{\boldsymbol{y}}_{b};\boldsymbol{\zeta}_{b})}{\partial \Delta\boldsymbol{\xi}_{b}[o_{3}]\partial\Delta\boldsymbol{\xi}_{b}[o_{3}]}=\frac{2}{\sigma_{b}^{2}}\mathrm{Re}\{\sum\nolimits_{m=1}^{M}\mathrm{Tr}(\sum\nolimits_{s=1}^{\bar{S}}f_{s}^{2}|\boldsymbol{s}_{b}(f_{s})|^{2}\nonumber\\
&\boldsymbol{H}(\tilde{\boldsymbol{r}}_{b^{\prime}},\tilde{\boldsymbol{t}}_{b})^{H}\boldsymbol{H}(\tilde{\boldsymbol{r}}_{b^{\prime}},\tilde{\boldsymbol{t}}_{b})\boldsymbol{w}\boldsymbol{w}^{H})\}. \label{proB13}    
\end{align}
\end{small}%
Finally, we focus on $\frac{\partial^{2}p(\Delta\boldsymbol{\xi}_{b})}{\partial\boldsymbol{\zeta}_{b}\partial\boldsymbol{\zeta}_{b}}$, and the extended expression of $\frac{\partial^{2}p(\Delta\boldsymbol{\xi}_{b})}{\partial\boldsymbol{\zeta}_{b}\partial\boldsymbol{\zeta}_{b}}$ is expressed as
\begin{small}
\begin{align}
\frac{\partial^{2}p(\Delta\boldsymbol{\xi}_{b})}{\partial\boldsymbol{\zeta}_{b}\partial\boldsymbol{\zeta}_{b}}=\left[\begin{matrix}
\frac{\partial^{2}p(\Delta\boldsymbol{\xi}_{b})}{\partial x_{T}\partial x_{T}}&\frac{\partial^{2}p(\Delta\boldsymbol{\xi}_{b})}{\partial x_{T}\partial x_{T}}&\frac{\partial^{2}p(\Delta\boldsymbol{\xi}_{b})}{\partial x_{T}\partial\Delta\boldsymbol{\xi}_{b}}\\
\frac{\partial^{2}p(\Delta\boldsymbol{\xi}_{b})}{\partial x_{T}\partial x_{T}}&\frac{\partial^{2}p(\Delta\boldsymbol{\xi}_{b})}{\partial x_{T}\partial x_{T}}&\frac{\partial^{2}p(\Delta\boldsymbol{\xi}_{b})}{\partial x_{T}\partial\Delta\boldsymbol{\xi}_{b}}\\
\frac{\partial^{2}p(\Delta\boldsymbol{\xi}_{b})}{\partial x_{T}\partial\Delta\boldsymbol{\xi}_{b}}&\frac{\partial^{2}p(\Delta\boldsymbol{\xi}_{b})}{\partial x_{T}\partial x_{T}}&\frac{\partial^{2}p(\Delta\boldsymbol{\xi}_{b})}{\partial \Delta\boldsymbol{\xi}_{b}\partial\Delta\boldsymbol{\xi}_{b}}
\end{matrix}\right].\label{proB14}
\end{align}
\end{small}%
Since $p(\Delta\boldsymbol{\xi}_{b})$ and $x_{T}$ and $y_{T}$ are uncorrelated, we have
\begin{small}
\begin{align}
&\frac{\partial^{2}p(\Delta\boldsymbol{\xi}_{b})}{\partial x_{T}\partial x_{T}}=0, \frac{\partial^{2}p(\Delta\boldsymbol{\xi}_{b})}{\partial y_{T}\partial y_{T}}=0, \frac{\partial^{2}p(\Delta\boldsymbol{\xi}_{b})}{\partial x_{T}\partial\Delta\boldsymbol{\xi}_{b}}=\boldsymbol{0}\nonumber\\
&\frac{\partial^{2}p(\Delta\boldsymbol{\xi}_{b})}{\partial y_{T}\partial x_{T}}=0, \frac{\partial^{2}p(\Delta\boldsymbol{\xi}_{b})}{\partial y_{T}\partial\Delta\boldsymbol{\xi}_{b}}=\boldsymbol{0}.\label{proB15}
\end{align}    
\end{small}%
Then, $\frac{\partial^{2}p(\Delta\boldsymbol{\xi}_{b})}{\partial \Delta\boldsymbol{\xi}_{b}\partial\Delta\boldsymbol{\xi}_{b}}$ is denoted as
\begin{small}
\begin{align}
\frac{\partial^{2}p(\Delta\boldsymbol{\xi}_{b})}{\partial \Delta\boldsymbol{\xi}_{b}\partial\Delta\boldsymbol{\xi}_{b}}=\frac{1}{\sigma_{b}^{2}}\frac{\partial \Delta\boldsymbol{\xi}_{b}^{T}}{\partial\Delta\boldsymbol{\xi}_{b}}\boldsymbol{\Sigma}_{\Delta}^{-1}\frac{\partial \Delta\boldsymbol{\xi}_{b}}{\partial\Delta\boldsymbol{\xi}_{b^{\prime}}}=\boldsymbol{I}.\label{proB16}
\end{align}       
\end{small}%
Thus, we have $\frac{\partial^{2}p(\Delta\boldsymbol{\xi}_{b})}{\partial \Delta\boldsymbol{\xi}_{b}\partial\Delta\boldsymbol{\xi}_{b}}=\boldsymbol{I}$.
\end{appendices}

\end{document}